%Paper: hep-th/9208035
%From: nathan@dirac.physics.sunysb.edu (Nathan Berkovits)
%Date: 12 Aug 1992 21:43:45 -0400

\magnification=1200
%
% NO NUMBER ON FIRST PAGE
%
\def\drho{{\partial _\rho}}

\def\drhobar{{\partial _{\bar \rho}}}
\def\dsr{{{{\partial^2 \rho}\over{\partial z^2}}}}
\def\half {{1 \over 2}}
\def\dz{{\partial _z}}
\def\dzbar{{\partial _{\bar z}}}
\def\dy{{\partial _y}}

\def\hdz{{\hat \partial_z}}
\def\plb{{+\bar l}}
\def\mlb{{-\bar l}}
\def\pmb{{+\bar m}}
\def\pnb{{+\bar n}}

\def\pl{{+ l}}
\def\ml{{- l}}
\def\mm{{- m}}
\def\hp{{h^+}}
\def\hm{{h^-}}
\def\hbp{{\bar h^+}}
\def\hbm{{\bar h^-}}
\def\sigl{{\sigma_l}}
\def\sigbl{{\bar\sigma_l}}
\def\sigm{{\sigma_m}}

\def\pp{{\kappa^+}}
\def\pd{{\kappa^-}}
\def\epb{{\bar \kappa^+}}
\def\ppb{{\bar \kappa^+}}
\def\pdb{{\bar \kappa^-}}
\def\emb{{\bar \kappa^-}}

\def\ep{{\kappa^+}}
\def\em{{\kappa^-}}
\def\mUj{{m^{U(1)}_j}}
\def\Dp{{D_+}}
\def\Dm{{D_-}}
\def\Dpb{{\bar D_+}}
\def\Dmb{{\bar D_-}}

\def\hDp{{\hat D_+}}
\def\hDpm{{\hat D_\pm}}
\def\hDm{{\hat D_-}}

\def\tgt{{\theta^{\alpha}\gamma^{\mu}_{\alpha\beta}\theta^{\beta}}}
\def\TgT{{\Theta^{\alpha}\gamma^{\mu}_{\alpha\beta}\Theta^{\beta}}}

\def\etp {{\eta^+}}
\def\etm {{\eta^-}}

\def\gmu {{\gamma^\mu_{\alpha\beta}}}
\def\lp {{\lambda^+}}
\def\lm {{\lambda^-}}

\def\lbp {{\bar\lambda^+}}
\def\lbm {{\bar\lambda^-}}
\def\wp {{w^+}}
\def\wm {{w^-}}
\def\wbp {{\bar w^+}}
\def\wbm {{\bar w^-}}
\def\vep {{\varepsilon^+}}
\def\vem {{\varepsilon^-}}

\def\vebp {{\bar\varepsilon^+}}
\def\vebm {{\bar\varepsilon^-}}
\def\hvep {{\hat\varepsilon^+}}
\def\hvem {{\hat\varepsilon^-}}
\def\hvepm {{\hat\varepsilon^\pm}}

\def\xp {{\xi^+}}
\def\xm {{\xi^-}}

\def\php {{\phi^+}}
\def\phm {{\phi^-}}
\def\phbp {{\bar\phi^+}}
\def\phbm {{\bar\phi^+}}
\def\sp {{\psi^+}}
\def\sm {{\psi^-}}
\def\sbp {{\bar\psi^+}}
\def\sbm {{\bar\psi^-}}
\def\dzxp {{\dz x^+ +\half\sp\dz\sm+\half\sm\dz\sp}}
\def\dzbarxp {{\dzbar x^+ +\half\sbp\dzbar\sbm+\half\sbm\dzbar\sbp}}
\def\xplb {{x^\plb}}
\def\xml {{x^\ml}}
\def\Gplb {{\Gamma^\plb}}
\def\Gml {{\Gamma^\ml}}
\def\Gbplb {{\bar\Gamma^\plb}}
\def\Gbml {{\bar\Gamma^\ml}}

\def\Gpmb {{\Gamma^\pmb}}
\def\Gpnb {{\Gamma^\pnb}}
\def\Gmm {{\Gamma^\mm}}

\def\bp {{\beta^+}}
\def\bm {{\beta^-}}
\def\gp {{\gamma^+}}
\def\gm {{\gamma^-}}
\def\bbp {{\bar\beta^+}}
\def\bbm {{\bar\beta^-}}

\def\tmlb {{\theta^{-\bar l}}}

\def\tbmlb {{\bar\theta^{-\bar l}}}
\def\Spl {{S^{+l}}}
\def\Smlb {{S^{-\bar l}}}
\def\Sbpl {{\bar S^{+l}}}
\def\Sbmlb {{\bar S^{-\bar l}}}

\def\Sbpm {{\bar S^{+m}}}

\tolerance=5000
\footline={\ifnum\pageno>1
       \hfil {\rm \folio} \hfil
    \else \hfil \fi}

\overfullrule=0pt %keeps overfull boxes from showing up in right margin
\baselineskip=18pt
\raggedbottom
\centerline{\bf Calculation of Green-Schwarz Superstring Amplitudes}
\centerline{\bf Using the N=2 Twistor-String Formalism}
\vskip 12pt
\centerline{Nathan Berkovits}
\vskip 12pt
\centerline{Address until Sept.92:}
\centerline{ITP, SUNY at Stony Brook, Stony Brook, NY 11794, USA}
\centerline{Address after Sept.92:}
\centerline{Math Dept., King's College, Strand, London, WC2R 2LS, United
Kingdom}
\centerline{e-mail: nathan@dirac.physics.sunysb.edu}
\centerline {ITP-SB-92-42}
\centerline {August 1992}
\vskip 24pt
\centerline {\bf Abstract}
The manifestly SU(4)xU(1) super-Poincar\'e invariant
free-field N=2 twistor-string action for the ten-dimensional
Green-Schwarz superstring is quantized using standard BRST methods. Unlike
the light-cone and semi-light-cone gauge-fixed Green-Schwarz actions, the
twistor-string action does not require interaction-point operators at the
zeroes of the light-cone momentum, $\dz x^+$, which complicated all previous
calculations.  After defining the vertex operator for the massless physical
supermultiplet, as well as two picture-changing operators and an instanton-
number-changing operator, scattering amplitudes for an arbitrary number of
loops and external massless states are explicitly calculated by evaluating
free-field correlation functions of these operators on N=2 super-Riemann
surfaces of the appropriate topology, and integrating over the global moduli.
Although there is no sum over spin structures, only discrete values of the
global U(1) moduli contribute to the amplitudes. Because the spacetime
supersymmetry generators do not contain ghost fields, the amplitudes are
manifestly spacetime-supersymmetric, there is no multiloop ambiguity, and
the non-renormalization theorem is easily proven. By choosing the
picture-changing operators to be located at the zeroes of $\dz x^+$,
these amplitudes are shown to agree with amplitudes obtained using the
manifestly unitary light-cone gauge formalism.
\noindent
\vfil\eject
\centerline{\bf I. Introduction}
\vskip 12pt
The calculation of scattering amplitudes using the Green-Schwarz
formulation of the ten-dimensional superstring is
manifestly spacetime supersymmetric, and therefore
contains advantages over analogous calculations using the
Neveu-Schwarz-Ramond formulation of the superstring. For example,
the calculation of scattering amplitudes for external fermions in the
Green-Schwarz formulation is no more difficult than the calculation for
external bosons, and the divergences that appear before summing over
spin structures in the NSR formulation of the superstring are
absent in the Green-Schwarz formulation.$^1$

Despite this motivation, almost all calculations of
superstring scattering amplitudes have been performed using the
Neveu-Schwarz-Ramond formulation. The reason is that after gauge-fixing
the N=1 worldsheet super-reparameterization invariances of the NSR
superstring, the covariant NSR action simplifies
to a quadratic free-field action, allowing amplitudes to be
calculated by evaluating free-field correlation functions on
N=1 super-Riemann surfaces.$^{2,3,4}$

In the Green-Schwarz formulation,
however, it has not been possible using the usual superspace
variables to gauge-fix the action to a free-field action. In both
the light-cone gauge$^{5,6,7}$ and semi-light-cone gauge,$^{8,9,10}$
the Green-Schwarz
action requires non-trivial interaction terms whenever $\dz x^+$=0
on the Riemann surface
(these zeroes of $\dz x^+ \equiv \dz x^0 + \dz x^9$
occur at $2g+N-2$ points for a g-loop N-string scattering
amplitude).\footnote\dag{
In reference 11, it is shown that by adding
a counterterm to the free semi-light-cone gauge-fixed Green-Schwarz
action, both conformal and Lorentz invariance can be preserved in the
effective action. However because their calculations are perturbative
around non-zero backgrounds for $\dz x^+$, they can not be used to
prove Lorentz invariance near $\dz x^+$=0. In fact, since the proposed
counterterm vanishes in the light-cone gauge, it is clear that the
semi-light-cone gauge-fixed action requires the same non-trivial
interaction term as the light-cone gauge-fixed action in order to
produce the correct scattering amplitudes.}
Because the locations of these interaction points are complicated
functions of the momenta of the external strings and of the modular
parameters of the Riemann surface, only tree and one-loop scattering
amplitudes involving four external massless states have been
expressed as Koba-Nielsen-like formulas using these methods.$^1$\footnote
\dag{
Restuccia and Taylor were able to analyze properties of the
multiloop Green-Schwarz scattering amplitudes, but only in certain
regions of moduli space.$^7$ Also, Mandelstam has proposed a Koba-Nielsen-like
formula for the tree-level scattering of N massless states, but he
did not derive it from a functional integral approach.$^{12}${\vskip 2pt}}
An additional problem caused by the non-trivial interaction terms is
that they must be defined in such a way that when two or more
interaction points approach each other, there are no short-distance
singularities. In practice, this requires introducing a contact-term
interaction into the light-cone Green-Schwarz action which further
complicates the analysis of scattering amplitudes.$^{13,14,7}$

The difficulty in gauge-fixing the Green-Schwarz covariant
action$^{15}$ to
a free-field action came from the lack of a geometrical interpretation
for the fermionic Siegel symmetries$^{16}$
of the Green-Schwarz superstring when
expressed in terms of the usual superspace variables. Unlike the N=1
superconformal invariance of the NSR superstring, there was little
understanding of the global moduli for these local Green-Schwarz
symmetries.$^{9}$

In a recent paper,$^{17}$ it was shown that by introducing new twistor-like
variables$^{18}$ into the Lorentz-covariant Green-Schwarz heterotic
superstring action, two of the eight fermionic Siegel-transformations
can be interpreted as worldsheet super-reparameterizations,$^{19,20}$ and
after gauge-fixing the N=2 worldsheet super-reparameterization
invariance and the remaining six Siegel symmetries, the Green-Schwarz
action simplifies to a free-field action on an N=2 super-Riemann surface.
This gauge-fixed action retains only a manifest SU(4)xU(1) subgroup of the
original SO(9,1) target-space super-Poincar\'e invariance, however it
is manifestly N=2 superconformally invariant on the worldsheet.
\footnote\ddag{
The idea of replacing Siegel-transformations with worldsheet
super-reparameterizations originated in the work of Sorokin, Tkach,
Volkov, and Zheltukhin on the superparticle,$^{21}$ although a connection
between N=2 worldsheet supersymmetry and spacetime supersymmetry had
already been found by other authors.$^{22}$ In fact it was even conjectured
that the sum over spin structures in the NSR formalism of the superstring
might be better understood as the U(1) moduli of an N=2 surface.$^{23}$}

Although the Lorentz-covariant N=2 twistor-string action is presently only
known for the heterotic Green-Schwarz superstring, it is easy to
generalize the gauge-fixed free-field action to non-heterotic versions of
the Green-Schwarz superstring. In this paper, the free-field action for the
Type IIB superstring is
quantized using the usual BRST methods, and after bosonizing some of
the matter and ghost fields, vertex operators are constructed for the
physical massless supermultiplet. Scattering amplitudes
with an arbitrary number of loops and external massless states are
explicitly calculated by evaluating free-field correlation functions
on an N=2 super-Riemann surface of the appropriate topology and
integrating over the global super-moduli of the surface.

It is easily
shown that these scattering amplitudes satisfy the non-renormalization
theorem, that is, all loop amplitudes with less than four external
massless states vanish.
By choosing light-cone moduli for the super-Riemann surface, it can
also be shown that these amplitudes
agree with amplitudes obtained using the Green-Schwarz
light-cone gauge formalism if one assumes a simple conjecture
concerning the
contribution of the contact-term interactions to the light-cone
gauge amplitudes. A proof of this conjecture would therefore
prove the unitarity of the twistor-string scattering amplitudes.

As in the NSR formalism,$^{3,4}$ it is convenient to perform the integration
over the anti-commuting moduli by introducing picture-changing operators.
Although the matter content of these operators resembles the matter content
of the light-cone interaction-point insertions, they differ
in the fact that their location on the surface does not affect the
scattering amplitudes (the ``multiloop
ambiguity'' will be discussed later in the
introduction). For this reason, amplitude calculations using the
twistor-string formalism do not require any knowledge about the
location of the string interaction points, and are therefore much
simpler than calculations using the light-cone gauge or semi-light-cone
gauge formalisms.

In addition, it is useful to introduce instanton-number-changing
operators in order to evaluate correlation functions on N=2
surfaces of non-zero U(1) instanton number (for
external states that transform in a given way under the U(1) subgroup
of the target-space SU(4)xU(1) invariance, only N=2 surfaces of a fixed
instanton number contribute to the scattering amplitude).
Like the picture-changing operators, the scattering amplitude is
independent of the location of the instanton-number-changing
operators.

Because the physical states of the Green-Schwarz string are manifestly
spacetime supersymmetric, there is no need to sum over spin structures
in order to project out unwanted states (in the twistor-string
scattering amplitudes, all spin structures contribute equally).
However the correlation
functions do depend on the global U(1) moduli of the N=2 surface,$^{24,25}$
and because of the presence of bosonized matter fields with negative
energies (these fields come from bosonizing bosons), these correlation
functions contain unwanted poles for all but special discrete values
of the U(1) moduli. Fortunately, the twistor-string formalism restricts
the region of integration for the global U(1) moduli to coincide with
these special values for which no unwanted poles occur.

It is well-known that bosonization of super-reparameterization ghosts
also introduces fields with negative energies, and therefore, correlation
functions with unwanted poles. Although the residues of these poles are
total derivatives in moduli space,$^4$ the presence of divergences in the
integrands of the scattering amplitudes would force the introduction
of cutoffs in the moduli space, possibly
creating surface term contributions.$^{26}$
In the NSR formalism, these cutoffs are necessary since before summing
over spin structures, the scattering amplitudes are not spacetime
supersymmetric and contain divergences.$^{27}$ The fact that the scattering
amplitude depends on the choice of the cutoff through the surface term
contributions is known as the ``multiloop ambiguity''. In the
twistor-string formalism, however, there is no multiloop ambiguity since
the amplitudes are manifestly spacetime supersymmetric. This fact is
easily demonstrated since unlike the spacetime supersymmetry generators
in the NSR formalism, the twistor-string spacetime supersymmetry
generators are independent of the bosonized ghost fields, and
therefore contain no unwanted poles.

Section II of this paper discusses quantization of the gauge-fixed N=2
twistor-string action in which the BRST charge is constructed, the
U(1)-transforming fields are bosonized, and the two picture-changing
operators, the instanton-number-changing operator,
and the massless physical vertex operators are defined.
Section III discusses the calculation of Green-Schwarz superstring
scattering amplitudes by describing tree amplitudes, beltrami differentials,
and multiloop correlation functions for the various matter and ghost fields.
Section IV analyzes the scattering amplitudes, showing that all spin
structures contribute equally, that
the non-renormalization theorem is satisfied, and that
the scattering amplitudes agree with amplitudes
obtained using the Green-Schwarz light-cone gauge formalism if one assumes
a simple conjecture concerning the contribution of the light-cone gauge
contact-term interactions.
Section V proposes applications for the results of this paper
and discusses possible approaches to Lorentz-covariantizing the
scattering amplitudes.
The Appendix reviews the gauge-fixing procedure for the
covariant N=(2,0) twistor-string action
of the Green-Schwarz heterotic superstring.
\vskip 24pt
\centerline {\bf II.Quantization of the N=2 Twistor-String}
\vskip 12 pt
A. The Gauge-Fixed Free-Field Action
\vskip 12 pt

It was recently shown that by introducing twistor-like
variables,$^{16}$ the Lorentz-covariant action for the ten-dimensional
Green-Schwarz heterotic superstring can be defined on an N=(2,0)
super-worldsheet.$^{19,20}$ These new variables allow two of the fermionic
Siegel-symmetries to be replaced with N=(2,0) super-reparameterizations,
and after gauge-fixing the super-reparameterizations and the remaining
six Siegel-symmetries, the Lorentz-covariant twistor-string action
reduces to a free-field action with manifest target-space SU(4)xU(1)
super-Poincar\'e invariance and manifest worldsheet N=(2,0)
superconformal invariance (this gauge-fixing procedure is reviewed
in the Appendix).$^{17}$
Unfortunately, at the present time there are no
Lorentz-covariant twistor-string actions for the non-heterotic
Green-Schwarz superstring. Nevertheless, it is straightforward to
generalize the gauge-fixed free-field action of equation (A.10) to the
Type IIB Green-Schwarz superstring by extending the N=(2,0)
super-worldsheet to an N=(2,2) super-worldsheet in the following way:
$$S=\int dz d\bar z d\pp d\pd d\ppb d\pdb [
X^{+\bar l} X^{-l}- W^-\Psi^+ -W^+\Psi^-
-\bar W^-\bar\Psi^+ -\bar W^+\bar\Psi^-]\eqno(II.1)$$
with the chirality constraints:
$$\Dm X^{+\bar l}=\Dmb X^{+\bar l}=
\Dp X^{-l}=\Dpb X^{-l}=0,\eqno(II.2)$$
$$\Dm\Psi^+=\Dmb\Psi^+=\Dp\Psi^-=\Dpb\Psi^-=
\Dm\bar\Psi^+=\Dmb\bar\Psi^+=\Dp\bar\Psi^-=\Dpb\bar\Psi^-=0,$$
$$\Dmb W^+=\Dpb W^-=\Dm\bar W^+=\Dp\bar W^-=0,$$
the N=(2,2) super-Virasoro constaints:
$$\Dp W^+ \Dm\Psi^- -\Dm W^- \Dp\Psi^++
\Dp X^{+\bar l}\Dm X^{-l}=0,\eqno(II.3)$$
$$\Dpb \bar W^+ \Dmb\bar\Psi^- -\Dmb \bar W^- \Dpb\bar\Psi^++
\Dpb X^{+\bar l}\Dmb X^{-l}=0 ;$$
and the non-local constraint:
$$\Omega\equiv\int_C dz d\pp d\pd |_{\bar \kappa^\pm=0} \Psi^-\Psi^+
+\int_C d\bar z d\ppb d\pdb |_{\kappa^\pm=0} \bar \Psi^-\bar\Psi^+ =0
\eqno(II.4)$$
where $C$ is any closed curve on the two-dimensional
surface (the non-local
constraint, $\Omega$, is the N=(2,2) version of equation (A.14),
and imposes restrictions on the U(1) moduli of the surface since
it implies that
$\int d\pp d\pd |_{\bar \kappa^\pm=0}
\Psi^-\Psi^+$ is a holomorphic one-form
with purely imaginary periods when integrated around a non-trivial loop).
Note that the N=(2,2) super-worldsheet has been Wick-rotated to Euclidean
space with coordinates $[z,\ep,\em;\bar z,\epb,\emb]$ satisfying
$\bar z=z^*, \epb=(\em)^*$, $\emb=(\ep)^*$ (the
action of equation (II.1) is
real since after Wick-rotation, the convention
$(\Phi_1 \Phi_2)^*=\Phi_1^* \Phi_2^*$ is used for both
bosons $and$ fermions);
$D_{\pm}\equiv \partial_{\kappa^\pm}+\half\kappa^\mp\partial_z$,
$\bar D_{\pm}\equiv \partial_{\bar\kappa^\pm}+\half\bar\kappa^\mp
\partial_{\bar z}$ (this definition
differs from that of the Appendix); and $X^{+\bar l}
=(X^{-l})^*$, $\bar W^\pm=(W^\mp)^*$, $\bar \Psi^\pm=(\Psi^\mp)^*$.
Under the 16 global spacetime-supersymmetry transformations
that preserve the gauge-fixing,
$$\delta X^{+\bar l}=\epsilon^{-\bar l}\Psi^++\bar\epsilon^{-\bar l}
\bar\Psi^+,
\delta X^{-l}=\epsilon^{+l}\Psi^-+\bar\epsilon^{+ l}\bar\Psi^-,
\eqno(II.5)$$
$$ \delta W^+=-
\epsilon^{+l} X^\plb, \delta W^-=-\epsilon^\mlb X^{-l}, \delta \bar W^+
=
-\bar\epsilon^{+l} X^\plb,
\delta \bar W^-=-\bar\epsilon^\mlb X^{-l},$$
and under the target-space SU(4)xU(1) rotations, $$[X^{+\bar l},
X^{-l}, \epsilon^{-\bar l}, \epsilon^{+l}, \bar\epsilon^\mlb,
\bar\epsilon^{+l},
W^\pm, \bar W^\pm, \Psi^\pm,\bar\Psi^\pm]$$ transforms
like a $[\bar 4_{+{1\over 2}}, 4_{-{1\over 2}},
\bar 4_{-{1\over 2}}, 4_{+{1\over 2}},
\bar 4_{-{1\over 2}}, 4_{+{1\over 2}},
1_{\pm 1}, 1_{\pm 1},
1_{\pm 1}, 1_{\pm 1}]$ representation (note that the SO(8)
anti-chiral spinor is chosen to break into a $[1_{+1},6_0,1_{-1}]$
representation of SU(4)xU(1), rather than the usual choice$^{5,6,7}$ of the
SO(8) vector).

After placing the auxiliary fields on-shell,
the free-field action of equation (II.1)
takes the following component form:
$$S=\int dz d\bar z (\dz x^\plb \dzbar x^\ml- \Gplb \dzbar\Gml
-\Gbplb \dz \Gbml \eqno(II.6)$$
$$-\wm\dzbar\lp-\vem\dzbar\sp-\wp\dzbar\lm-\vep\dzbar\sm-
\wbm\dz\lbp-\vebm\dz\sbp-\wbp\dz\lbm-\vebp\dz\sbm)$$
with the N=(2,2) super-Virasoro constraints:$^{28}$
$$\dz  x^\plb \dz \xml -\half (\Gplb \dz\Gml + \Gml\dz\Gplb)
\eqno(II.7)$$
$$ -\half(\wm\dz\lp -\lp\dz\wm)-\vem\dz\sp-
\half(\wp\dz\lm -\lm\dz\wp)-\vep\dz\sm=$$
$$\dz x^\plb \Gml + \vep\lm -\wm\dz\sp=
\dz x^\ml \Gplb + \vem\lp -\wp\dz\sm=
\Gplb\Gml+\wp\lm-\wm\lp=
0,$$
$$\dzbar  x^\plb \dzbar \xml -\half (\Gbplb \dzbar\Gbml + \Gbml\dzbar\Gbplb)
$$
$$-\half(\wbm\dzbar\lbp -\lbp\dzbar\wbm)-\vebm\dzbar\sbp-
\half(\wbp\dzbar\lbm -\lbm\dzbar\wbp)-\vebp\dzbar\sbm=$$
$$\dzbar x^\plb \Gbml + \vebp\lbm -\wbp\dzbar\sbm=
\dzbar x^\ml \Gbplb + \vebm\lbp -\wbm\dzbar\sbp=
\Gbplb\Gbml+\wbp\lbm-\wbm\lbp=
0,$$
and the non-local $\Omega$ constraint:
$$\int_C dz (\lm\lp-\half\sm\dz\sp-\half\sp\dz\sm)+
\int_C d\bar z (\lbm\lbp-\half\sbm\dzbar\sbp-\half\sbp\dzbar\sbm)=0,
\eqno(II.8)$$
where at $\kappa^\pm=\bar\kappa^\pm=0$,
$$X^{+\bar l}=\xplb\equiv
x^l+ix^{l+4}, D_+ X^{+\bar l}=\Gplb, \bar D_+ X^{+\bar l}=\Gbplb,
\eqno(II.9)$$
$$X^{-l}=\xml\equiv
x^l-ix^{l+4}, D_- X^{- l}=\Gml, \bar D_- X^{-l}=\Gbml, $$
$$D_\pm W^\pm =w^\pm,\quad D_\mp D_\pm W^\pm =\varepsilon^\pm, \quad
\Psi^\pm=\psi^\pm,\quad D_\pm\Psi^\pm=\lambda^\pm,$$
$$\bar D_\pm \bar W^\pm =\bar w^\pm,\quad\bar
 D_\mp \bar D_\pm \bar W^\pm =\bar\varepsilon^\pm, \quad
\bar\Psi^\pm=\bar\psi^\pm,\quad\bar D_\pm\bar\Psi^\pm=\bar\lambda^\pm.$$

For the rest of this paper, only the closed oriented chiral Green-Schwarz
superstring (type IIB) will be discussed, although it should be
straightforward to generalize the discussion
to the open and heterotic types (the Type IIA closed
string may present special problems since the two spacetime supersymmetries
transform differently under the SU(4)xU(1) subgroup). To conserve space,
most equations will be
written only for the right-handed sector of the Type IIB superstring,
and the corresponding
equations for the left-handed sector can be obtained by complex conjugation.
\vskip 12pt
B. Construction of the BRST Charge
\vskip 12pt

The action for the N=(2,2) super-Virasoro ghosts, $[B,C]$ and $[\bar B,
\bar C]$, is:$^{29}$
$$S_{ghost}= \int
dz d\bar z d\pp d\pd d\ppb d\pdb [BC+\bar B \bar C]\eqno(II.10)$$
with the chirality constraints,
$\bar D_- B =\bar D_+ C =0$,
and the super-Virasoro constraints,
$D_+ B D_- C +D_-B D_+C+\dz (BC)=0$.

Placing the auxiliary fields on-shell, this action in component form is:
$$S_{ghost}=-\int dz d\bar z (b \dzbar c + \beta^+\dzbar\gamma^- +
\beta^-\dzbar\gamma^+ + v\dzbar u
+\bar b \dz\bar c + \bar\beta^+\dz\bar\gamma^- +
\bar\beta^-\dz\bar\gamma^+ + \bar v\dz\bar u)$$
with the super-Virasoro constraints:
$$b\dz c +\beta^+\dz\gamma^-+\beta^-\dz\gamma^++v\dz u +\dz(bc+\half
(\beta^+\gamma^-+\beta^-\gamma^+))=\eqno(II.11)$$
$$(b+\half\dz v)\gamma^+ -\beta^+(u+\half\dz c)-\dz(v \gp+\bp c)=
(b-\half\dz v)\gamma^- +\beta^-(u-\half\dz c)+\dz(v \gm-\bm c)
=$$
$$\beta^+\gamma^-
-\beta^-\gamma^++\dz(cv)=0,$$
where at $\bar\kappa^\pm=0$,
$$B=v+\ep\bm-\em\bp+\ep\em b,\quad C=c+\ep\gm+\em\gp+\ep\em u.
\eqno(II.12)$$

Using the N=(2,2) super-Virasoro constraints of equation (II.11), a BRST
charge can be constructed in the following way:$^{29}$

$$Q=
\int dz d\pp d\pd |_{\bar \kappa^\pm=0}
[C(\Dp X^{+\bar l}\Dm X^{-l}+
\Dp W^+ \Dm\Psi^- -\Dm W^- \Dp\Psi^+)
\eqno(II.13)$$
$$+C\dz C B-D_+C D_-C B]
$$
$$
+\int d\bar z d\ppb d\pdb |_{\kappa^\pm=0}
[\bar C(
\Dpb X^{+\bar l}\Dmb X^{-l}+
\Dpb \bar W^+ \Dmb\bar\Psi^- -\Dmb \bar W^- \Dpb\bar\Psi^+)$$
$$+\bar C
\dzbar \bar C \bar B-\Dpb \bar C\Dmb \bar C\bar B]. $$

It is easy to check that $[Q,B]$ at $\bar\kappa^\pm=0$ is the sum of
the matter and ghost super-stress-energy tensors of equations
(II.7) and (II.11),
that $[Q,\Omega]=0$ where $\Omega$ is defined in equation (II.8),
and that $Q$ is nilpotent including normal-ordering effects,
since the central charge contribution of the matter fields is
$(4\times 2)+(4\times 1)-(2
\times 2)-(2\times 1)=+6$, while the contribution of the N=2 ghost
fields is $-26+(2\times 11)-2=-6$.

Physical vertex operators, $V$, can now be
defined by the conditions
$[Q,V]=[\Omega,V]=0$ and $V\not=[Q,B]$ for any $B.$

\vskip 12pt
C. Bosonization of the U(1) Current
\vskip 12pt
In order to construct the physical vertex operator,
it is useful to first bosonize matter and ghost fields
that appear in the U(1) current,
$$J_{U(1)}=\Gplb\Gml+\wp\lm-\wm\lp+\bp\gm-\bm\gp+\dz(cv).
\eqno(II.14)$$
As in the NSR
formalism of the superstring, an unfortunate consequence of
bosonization is that the worldsheet superfields must be broken
into their individual components.
The super-reparameterization ghosts, $\beta^\pm$ and
$\gamma^\pm$, are bosonized in the following standard way:$^3$
$$\bp= e^{-\phm}\dz\xp,\bm= e^{-\php}\dz\xm,\eqno(II.15)$$
$$\gp=e^{\php}\etp,\gm=e^{\phm}\etm$$
where all expressions are normal-ordered,
and as $y\to z$, $\dy\phm (y) \dz\phm (z)$ and
$\dy\php (y) \dz\php (z)$
$\to -(y-z)^{-2}$ and therefore have negative
energies, $\etm (y) \dz\xp (z) \to  (y-z)^{-2}$,
$\etp (y) \dz\xm (z) \to  (y-z)^{-2}$, and all other
operator products are non-singular.

The $\Gplb$ and $\Gml$ matter fields are also bosonized in the
standard way$^3$ as:
$$\Gplb=e^\sigl ,\Gml=e^{-\sigl}\eqno(II.16)$$
where as $y\to z$, $\dy \sigl (y) \dz\sigma_m (z)\to \delta_{l,m}
(y-z)^{-2}$.

Finally, the bosonization of the $\lambda^\pm$
and $w^\pm$ matter fields is less straightforward,
but the following formulas can be shown to have the correct
operator-product expansions:
$$\lp=(\dzxp)
 e^\hp +e^{-\hm},\quad \lm=e^{-\hp}\eqno(II.17)$$
$$\wp=e^\hp (\dz \hp+\dz\hm+x^- (\dzxp))+x^- e^{-\hm},\quad
 \wm=x^- e^{-\hp},$$
where as $y\to z$, $\dy x^+ (y) \dz x^-(z)\to (y-z)^{-2}$,
$\dy \hp (y) \dz\hm(z) \to (y-z)^{-2}$,
and the $+$ and $-$ indices of $x^+$ and $x^-$
refer to the target-space
light-cone indices $x^9 \pm x^0$ (the SO(9,1) metric is
$[-+++++++++]$).
Since $\dy h^1(y) \dz h^1(z) \to (y-z)^{-2}$ and
$\dy h^2(y) \dz h^2(z)\to -(y-z)^{-2}$ where
$$h^+\equiv {1 \over \sqrt2}(h^1+h^2)\quad {\rm and} \quad
 h^-\equiv {1\over \sqrt2}(h^1-h^2),\eqno(II.18)$$
$h^1$ and $h^2$
describe two chiral bosons that take values on a circle of
radius $\sqrt2$, one with positive energy and the other with negative
energy.
Note that by shifting the scalar fields
$h^+$ and $h^-$ by a constant, the relative
coefficients of the two terms in $\lp$ and
$\wp$ can be changed without affecting the
operator-product expansions.

In order to give the correct conformal weights for the unbosonized
fields, the bosonized scalar fields $[\phi^\pm,\sigl,h^1,h^2,
x^\pm]$ must have screening charges $q= [+2,0,+\sqrt2,0,0]$
and the $[\xi^\pm,
\eta^\pm]$ fields must have conformal weight [0,1]. It is easy to
check using the formula $c=1\mp 3q^2$
that the total contribution to the central charge of the
unbosonized fields is equal to the total contribution of the
bosonized fields.

By defining $x^+$ and $x^-$ to be real quantities (i.e., the same
$x^+$ and $x^-$ appear in the bosonizations of $[\lambda^\pm,w^\pm]$
and $[\bar\lambda^\pm, \bar w^\pm]$), this bosonization would seem to
guarantee that the $\Omega$ constraint of equation (II.8)
is satisfied (note
that $e^{-\hm} (y) e^{-\hp}(z)\to 0$ as $y\to z$).
However, it was shown in reference 17 that the $\Omega$ constraint
restricts the global U(1) moduli of the N=2 surface since only for
certain special values of the U(1) moduli is it possible to find
holomorphic fields, $\lp$ and $\lm$, such that the real part of
$\int_C dz (\lm\lp-\half\sm\dz\sp-\half\sp\dz\sm)$ vanishes around all
non-contractible loops. As will be shown in Section III.C.,
the bosonization prescription of equation (II.17)
also imposes a restriction
on the U(1) moduli since correlation functions of the
$e^{\pm\hp}$ and $e^{\pm\hm}$ fields are only well-defined (i.e., do not
have unwanted poles) for special values of the U(1) moduli.

Because $\varepsilon^\pm$ and $\bar\varepsilon^\pm$
must commute with $\lambda^\pm$ and $\bar\lambda^\pm$, they should
commute with $\dzxp$ and $\dzbarxp$, but not with $x^+$. It is
therefore convenient to define new fields,
$$\hat\varepsilon^\pm\equiv\varepsilon^\pm -\half\dz x^- \psi^\pm
-x^-\dz \psi^\pm,\quad
\hat{\bar{\varepsilon}}^\pm\equiv\bar\varepsilon^\pm
-\half\dzbar x^- \bar\psi^\pm-x^-\dzbar\bar\psi^\pm,
\eqno(II.19)$$
which no longer commute with $\lambda^\pm$ and $\bar\lambda^\pm$,
but which do commute with $x^+$. When expressed in terms of the
bosonized fields and $\hat\varepsilon^\pm$, it is easy
to check that the BRST charge is
invariant under constant shifts of $x^+$ and $x^-$.
\vskip 12pt
D.Picture-Changing Operators
\vskip 12pt
As in the NSR formalism for the superstring, it is useful to define
operators that change the ghost number (or picture) of a physical
vertex operator (right and left-handed ghost number is defined as
$\int dz (cb+uv+\dz \php-\dz\phm)$ and
$\int d\bar z (\bar c\bar b+\bar u\bar v+\dzbar\phbp
-\dzbar\phbm)$).
These operators are constructed in the usual way$^3$ by
commuting the BRST charge, Q, with the $\xi^\pm$ fields that
appear in the bosonized $[\beta^\pm,\gamma^\pm]$ system of equation
(II.15):
$$Z^+\equiv[Q,\xp]=e^\phm [\dz \xml\Gplb+\vem\lp-\wp\dz\sm+
(b-\half\dz v)\gamma^+ -v\dz\gp+c\xp],
\eqno(II.20)$$
$$Z^-\equiv[Q,\xm]=e^\php [
\dz x^\plb \Gml + \vep\lm -\wm\dz\sp+
(b+\half\dz v)\gamma^- +v\dz\gm+ c\xm].$$

Like the N=1 case, $\dz Z^\pm$ is BRST-trivial so changing the
location of the picture-changing operators changes the integrand of
the scattering amplitude by a total derivative in the moduli space.
As was already mentioned in the introduction, these total derivatives
are harmless for the twistor-string because the integrands are
manifestly spacetime supersymmetric and therefore contain no
divergences. Note that unlike the N=1 case, there are no inverse
picture-changing operators in the N=2 cohomology since there are no
terms in $Z^\pm$ that are proportional to $e^{2\phi^\mp}$.

\vfill\eject
E. Instanton-Number-Changing Operators
\vskip 12pt

After bosonizing the fields that transform under the worldsheet
U(1)-transformations, the U(1) current of equation (II.14)
can be written as:
$$J_{U(1)}=\dz(\sum_{l=1}^4 \sigl - \hp+\hm-\php+\phm+cv).
\eqno(II.21)$$

Although $\int dz J_{U(1)}$ is not a well-defined operator since
it is only defined up to $2\pi$, the operator
$$I^n \equiv \exp[n
(\sum_{l=1}^4 \sigl - \hp+\hm-\php+\phm+cv)]
\eqno(II.22)$$
is a well-defined operator when $n$ is an integer, and $I$ will be
called the right-handed
instanton-number-changing operator. Since $J_{U(1)}=
[Q,v]$, $\dz I^n=n J_{U(1)} I^n=[Q,nvI^n]$, and therefore, $\dz I^n$
is BRST-trivial. So $I^n$ shares the property of the picture-changing
operators that it is in the BRST cohomology, but amplitudes do not
depend on its location. It is easy to check that evaluating a correlation
function on an N=2 surface with instanton number $[n,\bar n]$
(i.e., $n=\int dz d\bar z \dzbar A_z$ and
$\bar n=\int dz d\bar z \dz A_{\bar z}$, where $A_z$ and $A_{\bar z}$
are the two components of the U(1) gauge field) is equivalent to
evaluating the correlation function on an N=2 surface with vanishing
instanton number, but with an insertion of the operator
$I^n\bar I^{\bar n}$.

By adding $I^{m(g-1)}\bar I^{n(g-1)}$
to the background charge, it is easily seen
that ``twisting'' the conformal weights of the fields by redefining
the Virasoro generators $L\to L+{m\over 2}\dz J_{U(1)}$ and
$\bar L\to \bar L+{n\over 2}\dzbar\bar J_{U(1)}$, is equivalent
to integrating over a genus g surface with its instanton number shifted
by $[m(g-1),n (g-1)]$.

\vskip 12pt
F. Massless Physical Vertex Operators
\vskip 12pt

The massless supermultiplet for the closed oriented chiral superstring
consists of 256 physical states, half fermionic and half bosonic.
Using SU(4)xU(1) super-Poincar\'e invariant notation,$^5$ this supermultiplet,
$$G (x^\mu, \tmlb, \tbmlb)=g^{-,-}(x^\mu)+\tmlb g^{-l,-}(x^\mu)
 +\tbmlb g^{-,-l}(x^\mu)
+...,\eqno(II.23)$$
can be expressed as a function of the ten spacetime bosonic
coordinates and 8 of the 32 spacetime fermionic coordinates, where
$\partial_{x^\mu}\partial_{x_\mu} G=0$. Half of the 32 spacetime
supersymmetry transformations are realized linearly on G by
$$\Spl G=\partial_\tmlb G,\quad \Smlb G=-i\tmlb\partial_{x^+} G,\quad
\Sbpl G=\partial_\tbmlb G,\quad \Sbmlb G=-i\tbmlb\partial_{x^+} G.
\eqno(II.24)$$

It is convenient to choose to break the target-space SO(9,1)
Lorentz invariance down to SU(4)xU(1) in such a way that the SO(8)
vector breaks into $\bar 4_{+\half}$ and $4_{-\half}$ representations
of SU(4)xU(1), the chiral SO(8) spinor
breaks into $4_{+\half}$ and $\bar 4_{-\half}$ representations
of SU(4)xU(1), and the anti-chiral SO(8) spinor
breaks into $1_{+1}$, $6_0$, and $1_{-1}$ representations
of SU(4)xU(1). Although this choice of breaking SO(9,1) down to
SU(4)xU(1) is not the usual one$^{5,6,7}$ in which the SO(8) vector breaks
into
$1_{+1}$, $6_0$, and $1_{-1}$ representations
of SU(4)xU(1), it is related to the usual choice by SO(8) triality.

With this unconventional choice, the $\tmlb=\tbmlb=0$ component of
$G$, $g^{-,-}(x^\mu)$, is one component of a direct product of
two anti-chiral SO(8) spinors, rather than one component of a
direct product of two SO(8) vectors. The vertex operator for this
state with momentum $p^\mu$ and ghost number $(-2,-2)$ is:
$$V_{-,-}=\biggl|(p^+)^{-2}c\sp\sm \exp(-\hp -\php-2\phm)
\biggr|^2 \exp(
ip^\plb x^\ml +ip^\ml x^\plb+ip^- x^++ip^+x^-)\eqno(II.25)$$
where $p^\mu p_\mu=0$ and $p^+$ is assumed to be non-zero.

It is straightforward
to check that this state is in the BRST cohomology and
that if any of the $\psi$'s are removed, the state becomes BRST-trivial since
$$c\sp \exp(-\hp-\php-2\phm+ip^+ x^-)=[Q,\dz\xm \sm c\sp
\exp(-2\php-2\phm+ip^+ x^-)],$$
$$c\sm \exp(-\hp-\php-2\phm+ip^+ x^-)=[Q,(p^+)^{-1}\dz\xp \sp c \sm
\exp(-2\hp-\php-3\phm+ip^+ x^-)].$$

By attaching an arbitrary number of picture-changing operators
to $V_{-,-}$, other ``pictures'' for
the vertex operator
can be constructed.
For example, one picture for the vertex operator of $g^{-,-}(x^\mu)$,
with ghost number $(-1,-1)$ is:
$$ Z^+ \bar Z^+
V_{-,-} =
\biggl|(p^+)^{-1}c\sm\exp(-\php-\phm)
\biggr|^2
\exp(ip^\plb x^\ml +ip^\ml x^\plb+ip^- x^++ip^+x^-).
\eqno(II.26)$$
Pictures for the vertex operator of $g^{-,-}$ with lower ghost
number than $(-2,-2)$ can be obtained by starting from
$$W_{-,-}\equiv
\biggl|{(p^+)}^{-M-1}c\prod_{m=0}^{M-1}
\partial^m_z \sm
\partial^m_z \sp \exp(-\hp -M\php-(M+1)\phm)
\biggr|^2 \exp(ip^\mu x_\mu),\eqno(II.27)$$
which has ghost number $(-2M,-2M)$
and satisfies $V_{-,-}=|Z^+ Z^-|^{2(M-1)} W_{-,-}$.

The easiest way to obtain the vertex operators for the other states
in the massless supermultiplet, $G$, is to first construct the 16
spacetime supersymmetry generators of equation (II.24)
that act linearly on G. These
generators are constructed out of the twistor-string matter fields
as follows:
$$\Spl=\int dz (\Gml\lp-\dz\xml\sp)\eqno(II.28)$$
$$=\int dz ((\dzxp)\exp(-\sigl+\hp)
+\exp(-\sigl-\hm)-\dz\xml\sp),$$
$$\Smlb=\int dz (\Gplb\lm-\dz\xplb\sm)=\int dz (\exp(\sigl-\hp)-\dz
\xplb\sm),$$
$$\Sbpl=\int d\bar z (\Gbml\lbp-\dzbar\xml\sbp)$$
$$=\int d\bar z ((\dzbarxp)\exp(-\sigbl+\hbp)
+\exp(-\sigbl-\hbm)-\dzbar\xml\sbp),$$
$$\Sbmlb=\int d\bar z (\Gbplb\lbm-\dzbar\xplb\sbm)
=\int d\bar z (\exp(\sigbl-\hbp)-\dzbar
\xplb\sbm).$$
Note that these generators commute with the BRST charge and
have the anti-commutation relations,
$\lbrace\Smlb,S^{+m}\rbrace=2\delta_{l,m}\int dz \dz x^+$,
$\lbrace\Sbmlb,\bar S^{+m}\rbrace=2\delta_{l,m}\int d\bar z \dzbar x^+$.

Unlike the NSR case,$^3$ these spacetime supersymmetry generators
contain no ghost fields and therefore do not have unwanted poles
coming from correlation functions of the $\phi^\pm$ fields.$^4$  This
lack of unwanted poles for $S$ and $\bar S$
means that spacetime supersymmetry is manifest
and there is no multiloop ambiguity in the scattering amplitudes.

The vertex operator for the other states in $G$ can now be
constructed by commuting various combinations of $\Spl$ and $\Sbpl$
with $V_{-,-}$ (note that $g^{-,-}(x^\mu)$ is the lowest component of the
superfield $G$, so
$[\Smlb, V_{-,-}]=[\Sbmlb, V_{-,-}]=0$). Since $\Spl$ and $\Sbpl$ commute
with $Z^\pm$ and $\bar Z^\pm$, this construction can be carried
out in any picture of $V_{-,-}$.

The vertex operator for the supermultiplet $G(x^\mu, \tmlb,\tbmlb)$
with momentum $p^\mu$ ($p^\mu p_\mu=0$ and $p^+\not= 0$) is
therefore:
$$V_{G(x^\mu,\tmlb,\tbmlb)}=\exp (\tmlb\Spl+\tbmlb\Sbpl) V_{-,-}
\eqno(II.29)$$
where the contours of $\Spl$ and $\Sbpl$ surround the vertex operator
$V_{-,-}$.
The overall factor of $(p^+)^{-4}$ in $V_{-,-}$ can easily be checked
by calculating the vertex operator for the $(-l,-m)$
component of the graviton state in the
ghost-number $(1,1)$ picture and setting all fermion fields to zero:
$$V_{(-l,-m)}=
\Spl\Sbpm \biggl|(Z^+)^2 Z^-
\biggr|^2 V_{-,-} =\eqno(II.30)$$
$$ c\bar c (\dz x^\ml-{p^\ml \over p^+}\dz x^+)
(\dz x^\mm-{p^\mm \over p^+}\dz x^+)\exp(ip^\mu x_\mu).$$
The corresponding vertex operators for the other components of the
graviton when all fermion fields are set to zero
can be obtained by using the appropriate combinations of
picture-changing and instanton-number-changing operators, e.g.,
$$V_{(\plb, -m)}
=(p^+)^{-1}\epsilon_{ijkl} S^{+i} S^{+j} S^{+k} \Sbpm  I
(Z^-)^2 Z^+ \bar Z^- (\bar Z^+)^2 V_{-,-}=\eqno(II.31)$$
$$ c\bar c (\dz x^\plb-{p^\plb \over p^+}\dz x^+)
(\dz x^\mm-{p^\mm \over p^+}\dz x^+)\exp(ip^\mu x_\mu).$$

\vskip 24pt
\centerline {\bf III. Calculation of Green-Schwarz Scattering Amplitudes}
\vskip 12pt
A. Tree Amplitudes
\vskip 12 pt

Tree-level scattering amplitudes for N massless states of the
Green-Schwarz superstring are calculated by evaluating free-field
correlation functions on the sphere of the operators $V_G$, $Z^\pm$,
$\bar Z^\pm$, $I$, and $\bar I$. The locations of $N-3$ of the vertex
operators are integrated over the sphere, whereas the locations of the
other operators are arbitrary.

The number of $Z^\pm$ and $I$ operators that need to be inserted on
the sphere can be determined from the rule that
$$<\biggl|\exp(-2\php-2\phm-\sqrt2 h^1) \sm\sp c\dz c\dz\dz c
\biggr|^2 >_{sphere}=1.\eqno(III.1)$$
Note that the background charges for the fields $\xi^\pm$ and $u$
do not appear in the normalization rule. This does not violate BRST
invariance since after bosonization, the zero modes of these fields
do not appear in the BRST charge. Since $V_G\sim |\exp(-\php-2
\phm)|^2$, $Z^\pm\sim \exp(\phi^\mp)$, and $I\sim \exp(\phm -\php)$,
$$n_{Z+}=2N-2-n_I,\quad n_{Z-}=N-2+n_I.\eqno(III.2)$$

A final relation for $n_I$ can be derived from the fact that the
charge
$$K\equiv\int dz (:\vep\sm-\vem\sp+\wp\lm-\wm\lp:)=\int dz (:\hvep\sm
-\hvem\sp:+\dz\hm-\dz\hp)
\eqno(III.3)$$
commutes with the BRST charge and has the following commutation
relations with the other operators:
$$[K,V_G]=(\tmlb\partial_\tmlb -1)V_G,\quad
 [K,Z^\pm]=0,\quad [K,I]=-2I.
\eqno(III.4)$$
The fact that $K$ commutes with the background charge implies that
for the component of the scattering amplitude with $Y$ $\tmlb$'s
$(0\le Y\le 4N)$,
$$n_I=\half (Y-N),\eqno(III.5)$$
where the different $\tmlb$ components of the amplitude correspond to
the scattering of different components of G. Since $Y-N$ is even by
fermion number conservation, this equation always has a solution with
integer-valued $n_I$.

Using equations (III.2) and (III.5), one finds
$$n_{Z^+}=\half(5N-Y-4),\quad
n_{Z^-}=\half(N+Y-2),\quad n_I=\half(Y-N).\eqno(III.6)$$
Since $Z^\pm$ are the only operators with $\hat\varepsilon^\pm$,
there must be at least $N-1$ of each of these picture-changing
operators in order to cancel all but one of the $\sp\sm$ terms
that come from the $V_G$'s (one of the terms must remain to provide
the background charge). From equation (III.2) and the complex
conjugate equation, this implies
that $1\le n_I\le N-1$ and $1\le \bar n_{\bar I}
\le N-1$, and therefore, the only non-zero components in the
tree-level scattering amplitude have between
$(2+N)$ and $(3N-2)$ $\tmlb$'s
and between
$(2+N)$ and $(3N-2)$ $\tbmlb$'s.

So the tree-level scattering amplitude for N massless states is:
$$A_{sphere}=\eqno(III.7)$$
$$\prod_{r=4}^{N} \int dz_r d\bar z_r
<\hat V_{G,r} (z_r,\bar z_r) \prod_{s=1}^3 V_{G,s} (z_s,\bar z_s)
\biggl|\sum_{n=1}^{N-1} I^n (Z^+)^{2N-2-n}(Z^-)^{N-2+n}
\biggr|^2>_{sphere}$$
where $V_{G,r}\equiv c\bar c \hat V_{G,r}$,
and $A_{sphere}$ is a function of $p_r^\mu$, $\theta^\mlb_r$, and
$\bar\theta^\mlb_r$ for $\mu=0$ to $9$, $l=1$ to $4$, and $r=1$ to $N$.
Note that
since terms with a different number of instanton-number-changing operators
only contribute to components of $A_{sphere}$ with a different number
of $\tmlb$'s, the instanton contribution to the action of $e^{i\vartheta
n_I}$ ($\vartheta$ is the instanton ``theta-parameter'') can be cancelled
by shifting $\tmlb\to e^{-\half
i\vartheta}\tmlb$, or equivalently, by rotating by $e^{i\vartheta}$ the
U(1) subgroup of the manifest SU(4)xU(1) target-space
invariance.$^{30}$

After expressing the operators in equation (III.7) in terms of the
bosonized fields, the free-field correlation functions are easy
to evaluate using the following formula$^3$ for chiral bosons, $\phi$, of
screening charge q:
$$<\prod_{i=1}^n \exp(c_i \phi (z_i)) >_{sphere} =
\delta_{-q,\Sigma c_i} \quad\prod_{i<j}(z_i -z_j)^{\pm c_i c_j}
\eqno(III.8)$$
where the $+$ sign is taking for positive-energy chiral bosons
and the $-$ sign is taken for negative-energy chiral bosons.
The only exception to this formula is for the screening charge
of the $(\eta^\pm,\xi^\pm)$
and $(u,v)$ fields, which is taken to be zero.

For example, the tree-level scattering amplitude for three massless
states is calculated as follows:

$$A_{sphere}=<\prod_{s=1}^3 V_{G,s} (z_s,\bar z_s) \biggl|I (Z^+)^3 (Z^-)^2
+I^2 (Z^+)^2 (Z^-)^3
\biggr|^2>_{sphere}
\eqno(III.9)$$
$$=\prod_{\mu=0}^9\delta(\sum_{s=1}^3 p^\mu_s)
\biggl| \prod_{l=1}^4\delta(\sum_{s=1}^3 p^+_s \theta_s^\mlb)
[(p_1^+ p_2^+ p_3^+)^{-1} P^{-a}\Theta^{-\bar a}+
\epsilon^{abcd}P^{+\bar a}\Theta^{-\bar b}\Theta^{-\bar c}
\Theta^{-\bar d}]\biggr|^2$$
where $P^\mu\equiv p^+_1 p^\mu_2- p^+_2 p^\mu_1$ and
$\Theta^{-\bar a}\equiv (p^+_3)^{-1} (\theta_1^{-\bar a}-
\theta_2^{-\bar a})$. Note that $P^\mu$ and $\Theta^{-\bar a}$, when
multiplied by the delta-functions, are
invariant up to a sign when the labels of the strings are
interchanged.$^{12}$

\vskip 12pt
B.N=2 Super-Beltrami Differentials
\vskip 12pt

Because N=2 super-Riemann surfaces of non-zero genus are not all
conformally equivalent, beltrami differentials
need to be introduced
to distinguish the different surfaces.$^{24}$ For a surface of genus g
with N punctures and instanton number $[n,\bar n]$,
the complex beltrami differentials, $M_T^i$, describe shifts in the
bosonic Teichmuller parameters, $m^T_i$ for $i=1$ to
$3g-3+N$, the complex differentials, $M_{U(1)}^j$, describe shifts in
the bosonic U(1) moduli, $m^{U(1)}_j$ for $j=1$ to g,
the complex differentials, $M_+^k$, describe shifts in the fermionic
supermoduli, $m^+_k$ for $k=1$ to $2g-2+N-n$, and the complex
differentials, $M_-^l$, describe shifts in the fermionic
supermoduli, $m^-_l$ for $l=1$ to $2g-2+N+n$.
The contribution of the instanton number, n, to the relative
numbers of fermionic moduli can be understood from the fact
that shifting the conformal weights of the fields by
$L\to L+{n\over {2(g-1)}}\dz J_{U(1)}$ is equivalent to shifting the
instanton number of the surface by n.

Using the notation of reference 4, the g-loop scattering amplitude for
N massless states is:
$$A_g=\sum_{n=-\infty}^{\infty}
\sum_{\bar n=-\infty}^{\infty}
\biggl|\prod_{i=1}^{3g-3+N}\int dm^T_i
\prod_{j=1}^{g}\int dm^{U(1)}_j
\prod_{k=1}^{2g-2+N-n}\int dm^+_k
\prod_{l=1}^{2g-2+N+n}\int dm^-_l\biggr|^2
$$
$$
\prod_{J=1}^g\int d\rho_{a_J} d\rho_{b_J}
\int_R DX^{\plb}DX^\ml \biggl|DW^+ D\Psi^- DW^- D\Psi^+ DB DC
\biggr|^2 \eqno(III.10)$$
$$\biggl|\delta(<M_T^i|b>)\delta(<M_{U(1)}^j|v>)
\delta(<M_+^k|\bp>)\delta(<M_-^l|\bm>)\biggr|^2$$
$$\exp(\rho_{a_J}\Omega_{a_J}+
\rho_{b_J}\Omega_{b_J})
\exp(S_{matter}+S_{ghost}) \prod_{r=1}^N
V_{G,r} (z_r,\bar z_r)\bigl|Z^+(z_r)\bigr|^2,$$
where $R$ is an N=2 super-Riemann surface of genus g and
instanton number $[n,\bar n]$, $V_{G,r} |Z^+|^2$ is in the picture
with ghost number $(-1,-1)$,
$\rho_{a_J}$ and
$\rho_{b_J}$ are Lagrange multipliers
for the $\Omega$ constraint of equation (II.8) (for $\Omega_{a_J}$, the
loop $C$ goes around the $a_J$-cycle, while for $\Omega_{b_J}$, the
loop goes around the $b_J$-cycle),
$S_{matter}$ and
$S_{ghost}$ are defined in equations (II.1) and (II.10), and
the $\delta(<M|B>)$ terms come from the N=2 superconformal gauge-fixing.

As in the NSR formalism,$^4$ the integration over the anti-commuting
moduli, $m^+_k$ and $m^-_l$, can be easily performed if one
chooses the fermionic beltrami differentials, $M^k_+$ and $M^l_-$,
to have the form:
$$M^k_+(z)=\dzbar ({1\over {z-w^+_k}})=\delta (z-w^+_k),
\quad M^l_-(z)=\dzbar ({1\over {z-w^-_l}})=\delta (z-w^-_l)\eqno(III.11)$$
where $w^+_k$ and $w^-_l$ are independent of the N=2 super-moduli.
Since the only dependence on $m^+_k$ and $m^-_l$ comes from the
action,
$$S=S|_{m^\pm=\bar m^\pm=0}
+m^+_k <M_+^k|[Q,\bp]>
+\bar m^+_{\bar k} <\bar M_+^{\bar k}|[Q,\bbp]>\eqno(III.12)$$
$$+m^-_l <M_-^l|[Q,\bm]>
+\bar m^-_{\bar l} <\bar M_-^{\bar l}|[Q,\bbm]>,$$
integration over $m^\pm$ and $\bar m^\pm$, when combined
with the $|\delta (<M_\pm|\beta^\pm>)|^2$ factors, introduces
the picture-changing operator insertions,
$$\biggl|\prod_{k=1}^{2g-2+N-n} \delta (\bp (w^+_k)) [Q,\bp (w^+_k)]
\prod_{l=1}^{2g-2+N+n} \delta (\bm (w^-_l)) [Q,\bm (w^-_l)]\biggr|^2$$
$$=
\biggl|\prod_{k=1}^{2g-2+N-n} Z^+(w^+_k)
\prod_{l=1}^{2g-2+N+n} Z^-(w^-_l)\biggr|^2.\eqno(III.13)$$

It is convenient to choose N of the Teichmuller parameters to be
the locations, $z_r$, of the vertex operators. This implies that
the beltrami differentials, $M^i_T$ for $i=3g-2$ to $3g-3+N$,
are $\dzbar H(\epsilon -|z-z_r|)$ where $H$ is the Heavyside
step-function and $\epsilon$ is small. With this choice, the effect
of the $\biggl|\prod_{i=3g-2}^{3g-3+N} dm^T_i \delta(<M^i_T|b>
\biggr|^2$ term is
to replace the N vertex operators,
$V_{G,r}(z_r,\bar z_r)$, with $\int dz_r
d\bar z_r
\hat V_{G,r}(z_r,\bar z_r)$ where
$V_{G,r}\equiv c\bar c\hat V_{G,r}$.

For the $g$ U(1) moduli, it is convenient to choose
$$m^{U(1)}_j={1\over {2\pi}}\int_{b_j} dz A_z,\quad
\bar m^j_{U(1)}={1\over{2\pi}}\int_{b_j} d\bar z A_{\bar z},
\eqno(III.14)$$
where $\int_{a_j} dz A_z$ and $\int_{a_j} d\bar z A_{\bar z}$
have been gauge-fixed to zero using the N=2 superconformal
transformations ($A_z$ and $A_{\bar z}$ are the two components of
the U(1) gauge field  and $a_j, b_j$ are the 2g non-trivial loops
of the surface with intersection properties $a_i\cap b_j=\delta_{i,j}$).
The corresponding beltrami differentials are
$$M^{U(1)}_j (z)=\dzbar \int_{a_j} dy_j ({1\over {z-y_j}}),
\eqno(III.15)$$
so $<M^{U(1)}_j|v>=
\int_{a_j} dy_j v(y_j)$.

As in the tree-level amplitudes, conservation of the charge
$K\equiv\int dz (:\hvep\sm
-\hvem\sp:+\dz\hm-\dz\hp)$ of equation (III.3) implies that the instanton
number $n$ must equal $\half(Y-N)$ for the component of the
scattering amplitude with $Y$ $\tmlb$'s. Furthermore,
only the picture-changing operators $Z^\pm$ contain
$\hat\varepsilon^\pm$, and since
$\hat\varepsilon^\pm$ contains $g-1$ more zero modes than
$\psi^\pm$,
there must be at least ($N+g-1$) $Z^+$'s and ($N+g-1$) $Z^-$'s
(each $V_{G,r}$ contains one $\sm$ and one $\sp$). Since there
are $(2N+2g-2-n)$ $Z^+$'s and $(N+2g-2+n)$ $Z^-$'s, this implies that
$1-g\le n\le N-1+g$, and therefore, only
components of the amplitude with between ($2-2g+N$) and ($2g-2+3N$)
$\tmlb$'s are non-zero.

Using this information, the g-loop amplitude, $A_g$, can be
written as:

$$A_g=
\biggl|\prod_{i=1}^{3g-3}\int dm^T_i
\prod_{j=1}^{g}\int dm^{U(1)}_j \biggr|^2
\int_R DX^{\plb}DX^\ml \biggl|DW^+ D\Psi^- DW^- D\Psi^+ DB DC \biggr|^2 $$
$$\prod_{J=1}^g\int d\rho_{a_J} d\rho_{b_J}
\biggl|\delta(<M_T^i|b>)
\int_{a_j} dy_j v(y_j)
\sum_{n=1-g}^{N-1+g}I^n (Z^+)^{2g-2+2N-n} (Z^-)^{2g-2+N+n}
\biggr|^2
\eqno(III.16)$$
$$\exp(\rho_{a_J}\Omega_{a_J}+
\rho_{b_J}\Omega_{b_J})
\exp(S_{matter}+S_{ghost}) \prod_{r=1}^N
\int dz_r d\bar z_r \hat V_{G,r} (z_r,\bar z_r),$$
where the locations of the $I$'s and $Z^\pm$'s are arbitrary
(changing their locations changes the integrand by a total
derivative in the moduli).

The above functional integral can be explicitly evaluated once
the free-field correlation functions on the surface, R,
for the matter and ghost fields have been determined.

\vfill\eject
C. Correlation Functions for the $W^\pm$ and $\Psi^\pm$ Superfields
\vskip 12pt

After bosonizing the $[\lambda^\pm,w^\pm]$ fields as in equation (II.17),
the $\Omega$ constraint,
$\int_C dz (\lm\lp-\half\sm\dz\sp-\half\sp\dz\sm)+
\int_C d\bar z (\lbm\lbp-\half\sbm\dzbar\sbp-\half\sbp\dzbar\sbm)=0$,
is trivially solved. However, the restriction on the U(1) moduli
that is imposed by this constraint does not disappear. Since the
bosonized fields include fields with negative energy, demanding
analyticity of these fields (i.e., that their correlation functions
do not have unwanted poles) will impose a similar
restriction on the U(1) moduli.

By expressing the unbosonized fields, [$\lambda^\pm,w^\pm, \psi^\pm,
\varepsilon^\pm$], in terms of the bosonized fields, [$x^\pm,h^1,h^2,
\psi^\pm,\hvepm$], all correlation functions can be performed using
the results of references 31 and 32 (because the zero-mode of
$x^+$ is a well-defined field on the surface, there is no analog
for the special treatment needed to handle the $\xi^\pm$ zero-mode
in the bosonization of the $[\beta^\pm,\gamma^\pm]$ system$^4$).
The anomolous contributions to
these correlation functions from the conformal factor can be safely
ignored since vanishing of the total central charge implies that
these anomolous contributions will cancel out in the final scattering
amplitude.

The $[\psi^+,\hvem]$ fields can be represented
by chiral scalar bosons with screening charge $q=1$ which take
values on a circle of radius 1, so their correlation functions are:$^4$
$$<\prod_{i=1}^m \hvem (y_i)\prod_{j=1}^n \sp (z_j)>_{\tau} =
Z([\sum_{i=1}^m y_i -\sum_{j=1}^n z_j - \Delta],\tau)
\eqno(III.17)$$
where $\Delta$ is the Riemann class,
$\tau$ is the period matrix of the surface and is a complex
symmetric $g\times g$ matrix with positive-definite
imaginary part,
$$Z([\sum_{i=1}^n c_i z_i -q \Delta],\tau)
=\delta_{q(g-1),\Sigma c_i}
\prod_{i<j} E(z_i, z_j)^{c_i c_j} \prod_{i=1}^n
\sigma (z_i)^{q c_i}
(Z_1(\tau))^{-\half}
\Theta([\sum_{i=1}^n c_i z_i -q\Delta],\tau),\eqno(III.18)$$
$Z_1(\tau)$ is a normalization for $Z$ such that
$Z([\sum_{i=1}^g z_i -y- \Delta],\tau)=Z_1(\tau)
\det_{ij} \omega_i (z_j)$, $\omega_i$ are the g canonical holomorphic
one-forms satisfying $\int_{a_j}\omega_i=\delta_{i,j}$ and
$\int_{b_j}\omega_i=\tau_{ij}$, $E(y,z)$ is the prime form,
$${{\sigma (y)}\over {\sigma (z)}}=
{{\Theta([y-\sum_{i=1}^g p_i +\Delta],\tau)}\over
{\Theta([z-\sum_{i=1}^g p_i +\Delta],\tau)}}
\prod_{j=1}^g {{E(y,p_j)}\over{E(z,p_j)}}\eqno(III.19)$$
for arbitrary $p_i$
(the final amplitudes will contain equal powers of $\sigma$ in the
numerator and denominator because of the vanishing conformal anomaly),
$[\sum_{i=1}^n (y_i -z_i)]_j\equiv\sum_{i=1}^n \int_{z_i}^{y_i}
\omega_j$ is an element in the Jacobian variety $C^g/ (Z^g+\tau Z^g)$,
and $\Theta(z,\tau)\equiv
\sum_{n\in Z^g} \exp(i\pi n_j\tau_{jk} n_k+2\pi in_j
z_j)$ which satisfies
$\Theta(z+\tau n+m,\tau)=
\exp(-i\pi n_j\tau_{jk} n_k-2\pi in_j
z_j)
\Theta(z,\tau)$ for $z\in C^g$ and $n,m\in Z^g$.
For a brief but sufficient review of these objects, see Chapter 3
of reference 31.

The $x^+$ and $x^-$ fields are non-chiral
scalar bosons which take values on the
real line, so their correlation functions are:

$$<\prod_{j=1}^n \exp(ip_j^+ x^- (z^j)+ip^-_j x^+(z_j))>_{\tau}
\eqno(III.20)$$
$$
=\delta (\sum_{j=1}^n p_j^+)
\delta (\sum_{j=1}^n p_j^-)
(\det Im\, \tau)^{-1} |Z_1(\tau)|^{-2}
\prod_{j\not= k} F(z_j,z_k)^{p^+_j p^-_k}
,$$
where $F(y,z)=\exp(-2\pi Im[y-z](Im\, \tau)^{-1} Im[y-z]) |E(y,z)|^2$.
Note that this non-holomorphic correlation function can be expressed
as:$^{33}$
$$\delta (\sum_{j=1}^n p_j^+)
\delta (\sum_{j=1}^n p_j^-)
\prod_{J=1}^g
\int_{-\infty}^{\infty} dk^+_J dk^-_J\eqno(III.21)$$
$$
\biggl| <\exp[\int_{b_J}dy_J(ik^+_J \partial_{y_J}
 x_z^- (y_J)+ik^-_J \partial_{y_J} x_z^+(y_J))]
\prod_{j=1}^n \exp(ip_j^+ x_z^- (z_j)+ip^-_j x_z^+(z_j))>_{\tau}\biggr|^2$$
where
$$<\prod_{j=1}^n \exp(ip_j^+ x_z^- (y_j)+ip^-_j x_z^+(y_j))>_{\tau}
\equiv(Z_1(\tau))^{-1}\prod_{j\not= k}E(y_j,y_k)^{p_j^+ p_k^-},$$
and $(k_J^\pm)^*\equiv -k_J^\pm$,
$(p_j^\pm)^*\equiv -p_j^\pm$ in the above formula. The integrations
over $k^+_J$ and $k^-_J$ in the above formula bear a close resemblance
to the Lagrange multipliers, $\rho_{a_J}$ and $\rho_{b_J}$, of
equation (III.10).

Since the chiral scalar bosons, $h^1$ and $h^2$, take values on a circle
of radius $\sqrt2$, their correlation functions are
not as straightforward to calculate. Nevertheless, these functions can be
determined using the results of reference 32. The $h^1$ fields
have screening charge $q=\sqrt2$, and so their correlation functions
are:

$$<\prod_{i=1}^n \exp(c^{1}_i h^1 (z_i))>_{\tau} =
f_{1}(\tau)
\delta_{\sqrt2 (g-1),\Sigma c^{1}_i}
\prod_{i<j} E(z_i, z_j)^{c^{1}_i c^{1}_j} \prod_{i=1}^n
\sigma (z_i)^{\sqrt2 c^{1}_i} \eqno(III.22)$$
$$
\Theta([\sum_{i=1}^n \sqrt2 c^{1}_i z_i -2\Delta],2\tau),$$
where $f_1(\tau)$ is an unknown overall normalization factor.
These correlation functions
for chiral bosons that take values on a circle of radius
$\sqrt2$ have the strange property that they are periodic when any of
the operators is taken once around an $a_j$-cycle, or $twice$
around a $b_j$-cycle.

The $h^2$ fields also take values on a circle of radius $\sqrt2$,
but differ from the $h^1$ fields in three ways. Firstly, the
$h^2$ field undergoes a shift of $2\sqrt2\pi i m^{U(1)}_j$ when it goes
around a $b_j$-cycle, and therefore a shift of $4\sqrt2\pi i m^{U(1)}_j$
when it goes twice around. Secondly, the energy of $h^2$ is negative,
so the correlation functions are inverted. And thirdly, $h^2$
has no screening charge. The correlation functions for the
$h^2$ fields are therefore:

$$<\prod_{i=1}^n \exp(c^{2}_i h^2 (z_i))>_{\tau} =
f_{2}(\tau)
\delta_{0,\Sigma c^{2}_i}
\biggl[
\prod_{i<j} E(z_i, z_j)^{c^{2}_i c^{2}_j}
\Theta([\sum_{i=1}^n \sqrt2 c^{2}_i z_i] -2m^{U(1)},
2\tau)\biggr]^{-1}.\eqno(III.23)$$

Putting together the correlation functions of equations (III.22) and
(III.23),
one finds:

$$<\prod_{i=1}^n \exp(c_i^{-} h^+ (z_i)+c_i^{+} h^- (z_i))>_{\tau}
=
f_1(\tau) f_2(\tau)
\delta_{g-1,\Sigma c^{-}_i}
\delta_{g-1,\Sigma c^{+}_i}
\eqno(III.24)$$
$$
\prod_{i\not= j} E(z_i,z_j)^{c^{-}_i c^{+}_j}
\prod_{i=1}^n
\sigma (z_i)^{c^{-}_i + c^{+}_i}
{{\Theta([\sum_{i=1}^n \sqrt2 c^1_i z_i -2\Delta],2\tau)}\over
{\Theta([\sum_{i=1}^n \sqrt2 c^2_i z_i ]-2m^{U(1)}
,2\tau)}},$$
where $h^\pm={1\over \sqrt2} (h^1\pm h^2)$ and
where $c_i^\pm={1\over \sqrt2} (c_i^1\pm c_i^2)$.

Since
$\Theta([\sum_{i=1}^n \sqrt2 c^2_i z_i ]-2 m^{U(1)},2\tau)$
has zeroes as a function of $z_i$, the correlation functions of the
$\hp$ and $\hm$ fields will have unwanted poles unless the zeroes of
$\Theta([\sum_{i=1}^n \sqrt2 c^1_i z_i -2\Delta],2\tau)$ occur at
the same locations as the zeroes of
$\Theta([\sum_{i=1}^n \sqrt2 c^2_i z_i ]-2 m^{U(1)},2\tau)$.
This is possible only if
$$\mUj=[{1\over\sqrt2}\sum_{i=1}^n (c^2_i\pm c^1_i)\mp \Delta]_j
+\alpha_j,\eqno(III.25)$$
where $\alpha_j$ is $0$ or $\half$, and $\mUj$
is defined such that $0\le Re(\mUj)<1$ and $0\le Im(\mUj)
(Im\, \tau)_{jk}^{-1}<1$ (this definition chooses one point in
the Jacobian variety, $C^g/(Z^g+\tau Z^g)$, however as will be shown
in Section IV.A., the total scattering amplitude is independent of this
choice).

The ambiguity in $\mUj$ comes from the fact that $\Theta(z,2\tau)=
\Theta(-z,2\tau)$ and $\Theta(z+1,2\tau)=\Theta(z,2\tau)$, and can
be fixed by analyzing the following correlation function:

$$F(y^+,y^-)
\equiv<\prod_{i=1}^n \exp(c_i^{-} h^+ (z_i)+c_i^{+} h^- (z_i))
\lp (y^+) \lm (y^-)>_{\tau} \eqno(III.26)$$ where
$\Sigma_{i=1}^n c^{-}_i=
\Sigma_{i=1}^n c^{+}_i=g-1$. From equations (II.17) and (III.24),
$$F(y^+,y^-)=
<\prod_{i=1}^n \exp(c_i^{-} h^+ (z_i)+c_i^{+} h^- (z_i))$$
$$
(\dzxp) (y^+) \exp(\hp(y^+)-\hp(y^-))>_{\tau} $$
$$=C(\dzxp)(y^+) \prod_{i=1}^n {{E(y^+,z_i)^{c_i^{+}}\sigma(y^+)}
\over{ E(y^-,z_i)^{c_i^{+}}\sigma(y^-)}} \eqno(III.27)$$
where $C$ is independent of $y^\pm$,
and therefore,
$F(y^+,y^-)\to \exp(2\pi i [\sum_{i=1}^n c^{+}_i -\Delta]_j)
F(y^+,y^-)$ when $y^+$ goes around the $b_j$-cycle. Since $\lp(y^+)\to
\exp(2\pi i\mUj)\lp(y^+)$ when $y^+$ goes around the $b_j$-cycle, the
correct choice for the U(1) moduli is:
$$\mUj =
[\sum_{i=1}^n c^{+}_i -\Delta]_j .\eqno(III.28)$$

So finally, the correlation function for the $h^\pm$ fields is
given by:
$$<\prod_{i=1}^n \exp(c_i^{-} h^+ (z_i)+c_i^{+} h^- (z_i))>_{\tau}
=\eqno(III.29)$$
$$N(\tau)\delta_{g-1,\Sigma c^{-}_i}
\delta_{g-1,\Sigma c^{+}_i}
\prod_{i\not= k} E(z_i,z_k)^{c^{-}_i c^{+}_k}
\prod_{i=1}^n \sigma(z_i)^{c_i^-+c_i^+}
\prod_{j=1}^g \delta (\mUj-
[\sum_{i=1}^n c^{+}_i -\Delta]_j) $$
where $N(\tau)$ is an overall measure factor that is independent of
the locations of the fields.

\vskip 12pt
D. Correlation Functions for $X$ Superfields
\vskip 12pt

The correlation functions for each pair of $\xplb$ and $\xml$
fields are the same as for the $x^\pm$ fields of equation (III.20),
that is:

$$<\prod_{j=1}^n \exp(ip_j^\plb x^\ml (z^j)+ip^\ml_j x^\plb(z_j))>_{\tau}
=\eqno(III.30)$$
$$\delta (\sum_{j=1}^n p_j^\plb)
\delta (\sum_{j=1}^n p_j^\ml)
(\det Im\, \tau)^{-1} |Z_1(\tau)|^{-2}
\prod_{j\not= k} F(z_j,z_k)^{p^\plb_j p^\ml_k}
,$$
where $F(y,z)=\exp(-2\pi Im[y-z](Im\, \tau)^{-1} Im[y-z]) |E(y,z)|^2$.

The correlation functions for the $\Gml$ and $\Gplb$ fields are
also straightforward, with the only subtlety coming from the
U(1) shift of $\exp(2\pi i \mUj)$ when $\Gplb$ goes around the
$b_j$-cycle. Since $\Gml$ and $\Gplb$ can be represented by
chiral bosons, $\sigl$, with no screening charge that take values on a
circle of radius 1, their correlation functions are:

$$<\prod_{i=1}^n \exp(c_i \sigl(z_i))>_{\tau} =
Z([\sum_{i=1}^n c_i z_i]-m^{U(1)} ,\tau),\eqno(III.31)$$
where $Z$ is defined in equation (III.18) and $\mUj=
[\sum_{i=1}^n c^{h^-}_i -\Delta]_j $.
There is no need to include other spin structures for the
theta-function since, as will be shown in Section IV.A., all spin
structures contribute equally to the total scattering amplitude.

\vskip 12pt
E. Correlation Functions for Ghosts
\vskip 12pt

The correlation functions for the $b$ and $c$ Virasoro ghosts
with screening charge $q=3$ is:

$$<\prod_{i=1}^m b(y_i)\prod_{j=1}^n c(z_j)>_{\tau} =
Z([\sum_{i=1}^m y_i -\sum_{j=1}^n z_j -3\Delta] ,\tau).\eqno(III.32)$$

For the $\beta^\pm$ and $\gamma^\pm$ fields, the only difference
with the NSR treatment of the bosonized super-reparameterization
ghosts comes from the contribution of the U(1) moduli (note
that in equation 36 of reference 4, a factor of $(Z_1)^{\half}$
was mistakenly omitted). Because the
zero mode of $\xp$ does not appear in any of the operators,
an extra field, $\xi^+(x_0)$, needs to be introduced into
correlation functions of the $\bp$ and $\gm$ fields. Since the
screening charge of the bosonized field, $\phm$, is +2, these
correlation functions are:$^4$

$$<\prod_{i=0}^p \xp(x_i)\prod_{j=1}^q \eta^-(y_j)
\prod_{k=1}^r \exp(c_k \phm(z_k))>_{\tau}
=\eqno(III.33)$$
$$\delta_{2(g-1),\Sigma c_k}\delta_{p,q}
{{\prod_{l=1}^q\Theta([-y_l+\sum_{i=0}^p x_i -\sum_{j=1}^q y_j +
\sum_{k=1}^r c_k z_k-2\Delta]+m^{U(1)} ,\tau)}\over
{\prod_{m=0}^p\Theta([-x_m+\sum_{i=0}^p x_i -\sum_{j=1}^q y_j +
\sum_{k=1}^r c_k z_k-2\Delta]+m^{U(1)} ,\tau)}}$$
$$
(Z_1)^{\half}
{{\prod_{i<i'} E(x_i,x_{i'})\prod_{j<j'} E(y_j,y_{j'})}\over
{\prod_{i<j} E(x_i,y_j)\prod_{k<k'} E(z_k,z_{k'})^{c_k
c_{k'}}
\prod_{k=1}^r \sigma(z_k)^{2c_k}}}.$$
Note that the correlation functions are independent of $x_0$ since
only the zero mode of $\xp(x_0)$ contributes.

Because the $u$ ghost does not appear in either the vertex operators,
the picture-changing operators, or the instanton-number-changing
operator, the correlation functions for the $u$ and $v$
ghosts must introduce an extra $u(x_0)$ field, just as the
$\bp$ and $\gm$ correlation functions required an extra
$\xp(x_0)$ field. Also, since the U(1) beltrami differentials
of equation (III.15) already introduce $g$ $v$ fields, there can be no
further contributions of $v$ fields from the other operators
(there must be $g-1$ more $v$ fields than $u$ fields to get a
non-zero amplitude, since the screening charge is $+1$).

The relevant correlation function for the $u$ and $v$ ghosts,
using equation (III.17), is therefore:
$$<u(x_0) \prod_{i=1}^g \int_{a_i} dy_i v(y_i)>_{\tau}=
\prod_{i=1}^g \int_{a_i} dy_i
Z([\sum_{j=1}^g y_j -x_0-\Delta] ,\tau)\eqno(III.34)$$
$$=Z_1(\tau)\prod_{i=1}^g \int_{a_i} dy_i \det_{jk} \omega_j (y_k)
=Z_1(\tau)\quad{\rm since}
\quad \int_{a_i} dy_i \omega_j(y_i)=\delta_{ij}.$$

The overall measure factor, $N(\tau)$, from the $h^\pm$
correlation function of equation (III.29) can now be fixed by
requiring that correlation functions without any $\hm$ fields,
when integrated over the U(1) moduli, are normalized to one.
This normalization prescription
will be shown in Section IV.C. to give amplitudes
which agree with amplitudes obtained using the light-cone gauge
formalism, and therefore is the correct unitary prescription.

Since the background charge normally requires $(g-1)$ $\hm$ fields, these
correlation functions should be evaluated on surfaces with
their instanton number shifted by $(g-1)$ (recall that this
shifts the conformal weight of $\lm$ from $\half$ to
$0$, and therefore shifts the
screening charge of $h^2$ from $0$ to $\sqrt2$).
On such surfaces, the correlation function of equation (III.29)
in the absence of $\hm$ fields is simply $N(\tau)\prod_{j=1}^g
\delta (\mUj)$, where $\sum_{i=1}^n c^{h^+}_i$ is assumed to be equal
to $2(g-1)$. So integration over the global U(1) moduli of this correlation
function, when combined
with the gauge-fixing contribution coming from the U(1) ghosts, $u$
and $v$, gives $Z_1(\tau) N(\tau)$. Therefore, normalization to one
prescribes that
$$N(\tau)=[Z_1(\tau)]^{-1}.\eqno(III.35)$$
This normalization prescription is consistent with
the vanishing of the conformal anomaly since
after shifting the screening charge of $h^2$,
the contribution of the $h^1$ and $h^2$ fields to the central
charge is $+2$ (the partition fuction of a $c=1$ chiral boson is
$(Z_1)^{-\half}$).

\vskip 24pt
\centerline {\bf IV. Analysis of the Scattering Ampltudes}
\vskip 12pt

A. Equivalence of Different Spin Structures
\vskip 12pt
By expressing the vertex operators and picture-changing operators
in terms of the bosonized free-fields and using the
results of Sections III.C.,D., and E. for evaluating their correlation
functions, the scattering amplitude of
equation (III.16) can be calculated as follows:

$$A_g=
<\biggl|\prod_{i=1}^{3g-3}\int dm^T_i
\prod_{j=1}^{g}\int dm^{U(1)}_j
\delta(<M_T^i|b>)
\int_{a_j} dy_j v(y_j)\biggr|^2\eqno(IV.1)$$
$$\biggl|
\xp\xm u
\sum_{n=1-g}^{N-1+g}I^n (Z^+)^{2g-2+2N-n} (Z^-)^{2g-2+N+n}
\biggr|^2
\prod_{r=1}^N
\int dz_r d\bar z_r \hat V_{G,r} (z_r,\bar z_r)>_{\tau},$$
where the locations of the
$I$'s, $Z^\pm$'s, $\xi^\pm$, and $u$ are arbitrary.
Note that for each combination of $\hm$ fields occuring in the
scattering amplitude, only one value of $\mUj$ contributes.

Because all operators in the correlation function are U(1)
singlets,
$$\sum_{l=1}^4 c^{\sigl}_i +c^{\hp}_i-c^{\hm}_i+c^{\php}_i
-c^{\phm}_i =0\eqno(IV.2)$$
for each $z_i$ that appears in the scattering amplitude. This
property implies the cancellation of all terms involving $E(y,z_i)$
and $\sigma (y)$ where $y$ is the location of an instanton-number-changing
operator, $I(y)$. Furthermore, all theta-functions in the amplitude
are independent of $y$
since for U(1)-transforming fields, the contribution to
the argument of the theta function from the U(1) moduli is
$\mp y$ (see equation (III.28)) while the contribution to the argument
from the fields is $\pm y$.
Therefore, the integrand of $A_g$ is completely independent of the
locations of the instanton-number-changing operators (this differs
from the picture-changing operators, since only $A_g$, and not the
integrand of $A_g$, is independent of the locations of the
$Z^\pm$'s).

Another consequence of equation (IV.2) is that changing the spin structures
of the U(1)-transforming fields does not
affect the integrand of $A_g$. This fact is not surprising$^{34}$ since
changing the spin structures from $[0]$ to $[\alpha]$
is equivalent to changing the
gauge-fixing condition on the U(1) gauge field, $A_z$, to
$\int_{a_j} dz A_z=2\pi i\alpha_{a_j}$ and
$\int_{b_j} dz A_z=2\pi i(\mUj+\alpha_{b_j})$, where $\alpha\in
({Z\over 2})^{2g}$. This affects equation (III.28) for the U(1) moduli
since the theta-function for $h^2$ in equation (III.23) now carries
spin structure $[\alpha]$. It is straightforward to check that
the zeroes of this theta-function coincide with the zeroes
of the theta-function for the $h^1$ field if
$$\mUj=\sum_{i=1}^n c^{\hm}_i-\Delta +\tau_{jk}\alpha_{a_k}
+\alpha_{b_j}.\eqno(IV.3)$$ Using the relation
$$\Theta ([\alpha],z-\tau\alpha_a-\alpha_b,\tau)
=\exp(-\pi i\alpha_{a_j}\tau_{jk}\alpha_{a_k}+2\pi i\alpha_{a_j}
z_j)\Theta(z,\tau)\eqno(IV.4)$$ and
equation (IV.2),
one can show that all phase factors in the total scattering
amplitude cancel out (the $\tau$ dependent factor cancels since
the $\sigl$ correlation functions contribute
$\exp(-4\pi i\alpha_{a_j}\tau_{jk}\alpha_{a_k})$,
the $\php$ and $\phm$ correlation
functions contribute
$\exp(2\pi i\alpha_{a_j}\tau_{jk}\alpha_{a_k})$,
and the $h^2$ correlation function contributes
$\exp(2\pi i\alpha_{a_j}\tau_{jk}\alpha_{a_k})$).

Similarly, one can show using equation (IV.2) and the periodicity
properties of $\Theta (z,\tau)$ that shifting the U(1) moduli,
$\mUj\to\mUj+\tau_{jk} p_k+q_j$ for $p_j$ and $q_j$ $\in Z$, does not
affect the integrand of $A_g$, and it is therefore unnecessary to
choose a region in the Jacobian variety, $C^g/(Z^g+\tau Z^g)$, when
defining $\mUj$.$^{25}$

\vskip 12pt
B. Proof of the Non-Renormalization Theorem
\vskip 12pt

The non-renormalization theorem for the superstring states that
all loop amplitudes with three or less massless particles vanish.
In the Green-Schwarz light-cone gauge$^7$ and semi-light-cone gauge$^9$
formalisms, this theorem can only be proven by explicitly assuming
Lorentz covariance for the scattering amplitudes.
In the Neveu-Schwarz-Ramond
formalism for the superstring, proof of the non-renormalization
theorem is complicated by
the possible contribution of
surface-terms from cutoffs in the moduli
space (these cutoffs
are necessary since the NSR amplitudes diverge before
summing over spin structures, but were ignored in the
proof of reference 35).

Using the expression for $A_g$ in equation (IV.1),
it will now be shown
that $A_g$ vanishes for $g\ge 1$ when there are three external
massless states.
Since the spacetime supersymmetry generators of equation (II.28),
$\Spl$, are analytic everywhere on the surface except at the
locations of the vertex operators, $z_i$ for $i=1$ to $3$, the
$\Spl$'s that encircle $z_3$ in $V_{G,3} (z_3,\bar z_3)$ can be
pulled off until they encircle either $z_1$ or $z_2$. This implies
that $A_g$ can only contain terms proportional to
$(\theta^\mlb_1-\theta^\mlb_3)$ and $(\theta^\mlb_2-\theta^\mlb_3)$, and by
fermion number conservation it must contain a total odd amount of
these factors. Supposing that the components of $A_g$ that we are
examining
have an even number of $(\theta^\mlb_1-\theta^\mlb_3)$ factors and an odd
number of $(\theta^\mlb_2-\theta^\mlb_3)$ factors, choose all
the picture-changing operators, $Z^\pm$, and instanton-number-changing
operators, $I^n$, to be located at $z_1$.

For these components of $A_g$, the contributing
components of $\hat V_{G,2}(z_2,\bar z_2)$ are
$$\hat V_{-l,-}=\Spl \hat V_{-,-}=p_2^+ \exp(-\sigl+\hp)\hat V_{-,-}\quad
{\rm and}\eqno(IV.5)$$
$$\hat V_{\plb,-}=
\epsilon_{lmnq} S^{+m} S^{+n} S^{+q} V_{-,-}=
\epsilon_{lmnq}
(p_2^+)^3 \exp(-\sigm-\sigma_n-\sigma_q+3\hp)\hat V_{-,-}$$
$$+
\epsilon_{lmnq}
(p_2^+)^2 :\exp(-\sigm-\sigma_n-\sigma_q+2\hp-\hm)\hat V_{-,-}:,
\eqno(IV.6)$$
where the $\sp$ term in $\Spl$ does not contribute since $\hat V_{-,-}$
of equation (II.25)
is proportional to $\sp$. Note that the last term of
equation (IV.6) requires normal-ordering because of singularities between
$\hm$ and $\hp$.

For the component $\hat V_{-l,-}$, the correlation function of
equation (III.31) for the
$\Gpmb$ and $\Gmm$ fields where $m\not= l$ is proportional
to $\Theta ([(g-1)z_1 -\Delta],\tau)$ since all of the $\sigm$
and $\hm$ fields are located at $z_1$. But
this is zero by Riemann's vanishing theorem when $g\ge 1$, since
$\Theta ([\sum_{i=1}^{g-1} y_i -\Delta],\tau)=0$ for
arbitrary $y_i$. Similarly for the first term in the
component, $\hat V_{\plb,-}$, the correlation function for
the $\Gplb$ and $\Gml$ fields is proportional to
$\Theta ([(g-1)z_1 -\Delta],\tau)$, and is therefore zero.
For the second term in the component,
$\hat V_{\plb,-}$, the correlation function for
the $\Gpmb$ and $\Gmm$ fields where $m\not= l$ is proportional
to
$\Theta ([(g-1)z_1 -\Delta],\tau)=0$, since the argument of the
theta-function receives $-z_2$ from the $\sigm$
field and $+z_2$ from the $\hm$ field.

To prove the vanishing of $A_g$ for two massless states whose
vertex operators are located at $z_1$ and $z_2$, pull all of the
$\Spl$'s off of $z_2$ and encircle them around $z_1$, and then
place all of the picture-changing and instanton-number-changing
operators at $z_1$. Then the correlation function for the
$\Gplb$ and $\Gml$ fields is proportional to
$\Theta ([(g-1)z_1 -\Delta],\tau)=0$ for all $l$.
Finally, for one massless state or no states, choose the locations
of all of the picture-changing-operators and
instanton-number-changing operators to coincide
with the vertex operator (or at any point, $z_1$, if there are
no states). Then once again, the correlation function for the
$\Gplb$ and $\Gml$ fields is proportional to
$\Theta ([(g-1)z_1 -\Delta],\tau)=0$ for all $l$.

So the superstring amplitudes, $A_g$, that were calculated using
the twistor-string formalism of the Green-Schwarz superstring,
have been proven to satisfy the non-renormalization theorem.

\vskip 12pt
C. Agreement of $A_g$ with Light-Cone Gauge Amplitudes
\vskip 12pt

The light-cone gauge formalism for calculating Green-Schwarz
superstring scattering amplitudes was first developed by
Green, Schwarz,$^5$ and Mandelstam,$^6$ and more recently, by Restuccia
and Taylor.$^7$ As discussed in reference 17, the Type IIB light-cone
gauge action on a Wick-rotated two-dimensional worldsheet
parameterized by $\rho$ and $\bar \rho\equiv (\rho)^*$ is:
$$S_{LC}=\int d\rho
d\bar \rho (\drho x^i \drhobar x^i -s^\alpha \drhobar
s^\alpha-\bar s ^\alpha\drho \bar s^\alpha).\eqno(IV.7)$$
After breaking SO(8) down to SU(4)xU(1) in such a way that
the SO(8) vector, $x^i$, splits into $\bar 4_{+\half}$
and $4_{-\half}$ representations of SU(4)xU(1) while the
SO(8) chiral spinor, $s^\alpha$, splits into $4_{+\half}$
and $\bar 4_{-\half}$ representations of SU(4)xU(1), the light-cone
gauge interaction term (ignoring contact terms) is $|H^-(\tilde\rho_a)
+H^+(\tilde\rho_a)|^2$, where
$$H^-(\tilde\rho_a)\equiv \lim_{\rho\to\tilde\rho_a}
(\rho-\tilde\rho_a)(\drho x^\ml s^\mlb)=
(\dsr)^{-1}( \dz x^\ml
\hat s^\mlb )(\tilde z_a),\eqno(IV.8)$$
$$H^+(\tilde\rho_a)\equiv \lim_{\rho\to\tilde\rho_a}
(\rho-\tilde\rho_a)^2\epsilon^{klmn}
(\drho x^{+\bar k} s^\mlb s^{-\bar m} s^{-\bar n})
=(\dsr)^{-2}
\epsilon^{klmn}(\dz x^{+\bar k}\hat s^{-\bar l}
\hat s^{-\bar m} \hat s^{-\bar n} )
(\tilde z_a) ,\eqno(IV.9)$$
$$\hat s^{\mlb} (z)\equiv (\dz\rho) s^{\mlb} (\rho(z))
\quad \quad{\rm and}\quad\quad
\hat s^\pl (z)\equiv s^\pl (\rho(z))$$ are conformal weight
one and zero fields$^7$ as functions of $z$
(this implies that
$s^\mlb(\rho(z))$ has zeroes at the punctures,
$z_r$ for $r=1$ to $N$, and poles at the interaction
points, $\tilde z_a$ for $a=1$ to $2g-2+N$, whereas $s^\pl(\rho(z))$
is regular at these points),
$\tilde\rho_a\equiv\rho (\tilde z_a)$ are the $(2g-2+N)$ interaction
points where $\dz\rho (\tilde z_a)=0$, and $\rho(z)$ is the unique
meromorphic function that maps the $g$-loop string diagram onto a genus
$g$ surface with $N$ punctures such that $Re (\rho)$ is single-valued
and $\dz\rho$ has poles with residue $p^+_r$ at the points $z_r$
for $r=1$ to $N$.

If the contributions from the contact term interactions are ignored,
the $g$-loop light-cone gauge
scattering amplitude for $N$ massless states is:

$$A_g^{LC}=\int\prod_{a=1}^{2g-2} d\tilde\rho_a d\tilde{\bar
\rho_a} \prod_{I=1}^g d\alpha_I d\phi_I
\int Dx^\plb Dx^\ml |D\hat s^\mlb D\hat s^\pl|^2\eqno(IV.10)$$
$$\exp[\int dz d\bar z
(\dz x^\plb \dzbar x^\ml -\hat
 s^\pl \dzbar \hat s^\mlb -\hat{\bar s}^\pl\dz\hat{\bar s}^\mlb)]
\prod_{r=1}^n V_{G,r}^{LC} (z_r)$$
$$\biggl|\sum_{n=0}^{2g-2+N}\sum_{a_+,a_-\in
a}\prod_{a^+=1}^{n} H^+
(\tilde \rho_{a^+})
\prod_{a^-=1}^{2g-2+N-n}H^-(\tilde\rho_{a^-})\biggr|^2,$$
where
$\tilde\rho_a$, $ \tilde{\bar \rho_a}$,
$\alpha_I$, and $\phi_I$ are the $(6g-6+2N)$
real light-cone moduli of interaction-point locations,
internal $p^+$'s, and twists, $\sum_{a_+,a_-\in a}$ means that
the $2g-2+N$ interaction points, $\tilde \rho_a$, should be
split up into
two subsets, $\tilde\rho_{a^+}$ and
$\tilde\rho_{a^-}$, in all possible ways, and
$$V_{G,r}^{LC}(z_r)=
|(p^+_r)^{-1}\exp(p^+_r \theta_r^\mlb
\hat s^\pl (z_r))|^2\exp(ip_r^\ml x^\plb (z_r)
+ip_r^\plb x^\ml(z_r)+p_r^- Re[\rho(z_r)])\eqno(IV.11)$$
(the light-cone vertex operator, $V_G^{LC}$,
has been normalized such that it agrees with
the matter part of the
vertex operator, $V_G$, in the ghost-number zero picture
when the non-light-cone fields, $\psi^\pm$ and $h^\pm$
have been set to zero, and $\Gml$ has
been identified with $\hat s^\pl$).
Note that the locations of the $\tilde\rho_a$'s are determined by
the $p_r^+$'s and $z_r$'s ,
but in a very complicated way. It is this complicated
dependence that makes Green-Schwarz light-cone gauge amplitudes
difficult to explicitly evaluate.

The first step in comparing $A_g$ of equation (IV.1) with $A_g^{LC}$
of equation (IV.10) is to choose light-cone moduli for $A_g$
(these light-cone moduli depend not only on the surface, but
also on the $p^+_r$ momenta of the N vertex operators), and to
insert ($2g-2+N$) $Z^+$'s and $(2g-2+N)$ $Z^-$'s at the interaction
points of the string diagram where $\dz\rho=0$.

In order to
ensure that there are enough picture-changing operators available
to do this, the picture for the vertex operators should be chosen
to have ghost number $(-4,-4)$, so the vertex operators are:
$$W_{G,r}(z_r,\bar z_r)=\eqno(IV.12)$$
$$
|(p^+_r)^{-3}c\sm\dz\sm\sp\dz\sp\exp(-\hp-2\php-3\phm)
\exp(\theta_r^\mlb \Spl)|^2
\exp(ip^\mu_r x_\mu)(z_r,\bar z_r).$$ It is easy to check that
$V_{G,r}=|Z^+ Z^-|^2 W_{G,r}$, so the number of
available $Z^+$ picture-changing operators is
$2g-2+3N-n_I$ and the number of available
$Z^-$ picture-changing operators is
$2g-2+2N+n_I$.
Since the instanton number, $n_I$, satisfies $n_I=\half(Y-N)$
where $0\le Y\le 4N$, there are enough available picture-changing
operators to insert one of each type at all the interaction points.
The extra ($2N-n_I$) $Z^+$'s and ($N+n_I$) $Z^-$'s can be
inserted anywhere on the string diagram.

Since $A_g^{LC}$ ignores the contribution of the light-cone
contact-term interactions (these contact-terms are necessary in the
light-cone formalism in order
to cancel the non-Lorentz-invariant divergences that occur when two
interaction-points approach each other),$^{13,14,7}$
one should not expect that $A_g^{LC}$
is precisely equal to $A_g$. However, a simple conjecture is that
the contribution of the light-cone contact-term interactions
is equivalent to the contribution of the fields
$$e^{\phi^{\mp}} [-w^{\pm}\dz\sm+
(b\mp\half\dz v)\gamma^\pm \mp v\dz\gamma^\pm+c\xi^\pm]\eqno(IV.13)$$ in
the picture-changing operators $Z^\pm$ of equation (II.20), plus
the contribution from the moduli dependence of the interaction-point
locations (this moduli dependence of the locations
of the picture-changing operators
implies that the beltrami differentials for the teichmuller
parameters, $M_T^i$, depend on the fermionic moduli, $m_k^\pm$)$^4$.
Therefore ignoring the contact-term interactions in $A_g^{LC}$
is conjectured to be equivalent
to using the following truncated form of the
picture-changing operators of equation (II.20):
$$\hat Z^+\equiv
e^{\phm} [\dz x^\ml \Gplb +\hvem\lp]\quad {\rm and}\quad
\hat Z^-\equiv e^{\php} [\dz x^\plb \Gml +\hvep\lm],\eqno(IV.14)$$
and ignoring the dependence of $M_T^i$ on the fermionic moduli.

It is easily checked that an analogous conjecture for the
Neveu-Schwarz-Ramond string is correct. This conjecture states
that ignoring the contribution of the NSR light-cone contact-term
interactions is equivalent
to ignoring the dependence of $M_T^i$
on the NSR fermionic moduli and
using the truncated NSR picture-changing
operator, $\hat Z\equiv
e^\phi [\dz x^i \Gamma^i +\dz x^+ \Gamma^-]$ at the
interaction points ($i=1$ to $8$),
rather than the full BRST-invariant
operator, $Z=e^\phi \dz x^\mu \Gamma_\mu+e^{2\phi}\dz\eta b+
\dz(e^{2\phi} \eta b) +c\dz\xi$.

This NSR conjecture can be proven by not integrating out the
anti-commuting moduli in the BRST formalism, and comparing the
resulting amplitudes with
the supersheet formalism of the light-cone gauge amplitudes,$^{36}$ in
which the contact-term interactions are automatically included.
Since the BRST amplitudes coincide (without using the conjecture)$^{37}$
with amplitudes obtained from the light-cone supersheet
formalism, and also coincide (using the conjecture) with
amplitudes obtained from the ordinary component form of the
light-cone formalism,$^{38}$ the NSR conjecture must be valid.
If a similar proof could be found for the Green-Schwarz conjecture,
it would prove that the twistor-string amplitudes, $A_g$,
agree with amplitudes obtained using the Green-Schwarz
light-cone formalism, and that they are therefore unitary.

With this conjecture, the truncated twistor-string amplitude,
$\hat A_g$, is:
$$\hat A_g=
<\int
\prod_{a=1}^{2g-2} d\tilde\rho_a d\tilde{\bar
\rho_a} \prod_{I=1}^g d\alpha_I d\phi_I
\biggl|\prod_{j=1}^{g}\int dm^{U(1)}_j
\prod_{i=1}^{3g-3+N}\int
\delta(<M_T^i|b>)
\int_{a_j} dy_j v(y_j)\biggr|^2\eqno(IV.15)$$
$$\biggl|\prod_{a=1}^{2g-2+N} (\hat Z^-\hat Z^+)(\tilde
z_a)
\xp\xm u
\sum_{n_I=1-g}^{N-1+g}I^{n_I} (\hat Z^+)^{2N-n_I} (\hat Z^-)^{N+n_I}
\biggr|^2
\prod_{r=1}^N
W_{G,r} (z_r,\bar z_r)>_{\tau},$$
where $\hat Z^\pm$ is defined in equation (IV.14), and $W_{G,r}$
is defined in equation (IV.12).

Since the vertex operators $W_{G,r}$ contribute $2N$ $\sp$'s
and $2N$ $\sm$'s, there must be at least ($2N+g-1$) $\hvem$'s
and ($2N+g-1$) $\hvep$'s coming from the truncated picture-changing
operators, $\hat Z^+$ and $\hat Z^-$. This means that the
$\hat Z^+ \hat Z^- (\tilde z_a)$ factors at the interaction
points must contribute at least $(g-1+n_I)$ $\hvem$'s and
$(g-1+N-n_I)$ $\hvep$'s.
But since $x^-$ appears only at the vertex operators, $\dz x^+$ has
the same poles and residues as $\dz \rho$, implying that
$\dz x^+(\tilde z_a)=\dz\rho(\tilde z_a)=0$
in correlation functions of the $x^-$
and $x^+$ fields. Therefore, no factor of $\hat Z^+ \hat Z^-
(\tilde z_a)$
at the interaction
points can contribute $\hvem \hvep$, since any such term would be
proportional to $\lm \lp$
(recall that $\lm\lp=\dzxp$, and if more $\sp\sm$ terms are
introduced, one needs even more $\hvem\hvep$'s).

So the only way to have enough $\hvepm$'s
is if $(g-1+n_I)$ interaction-point factors
at $z=\tilde z_{a^+}$
contribute
$\hvem\lp\exp (\php+\phm)\dz x^\plb \Gml$, the other
$(g-1+N-n_I)$ interaction-point factors at $z=
\tilde z_{a^-}$ contribute
$\hvep\lm\exp (\php+\phm)\dz x^\ml \Gplb$, and the
remaining $(2N-n_I)$ $\hat Z^+$'s and $(N+n_I)$ $\hat Z_-$'s
contribute $\hvem\lp\exp(\phm)$ and
$\hvep\lm\exp(\php)$.

In order to compare with $A_g^{LC}$, it is convenient to
insert one instanton-number-changing operator, $I(z)$, at each of the
$(g-1+n_I)$ interaction-points, $z=\tilde z_{a^+}$ (recall that
none of the correlation functions depend on the locations of
the instanton-number-changing operators, so there is no problem
with making the locations of the $I(z)$'s change with different
choices for the $\tilde z_{a^+}$'s). Since this inserts ($g-1$)
more instanton-number-changing operators than is necessary, it
shifts the conformal weights of the U(1)-transforming fields
by $\pm\half$ ($[\Gml,\Gplb,\lm,\lp,\wm,\wp,\gm,\gp,\bm,\bp]$
now have conformal weights $[0,1,0,1,0,1,-1,0,1,2]$).

After making these modifications and using equations (II.22) and (IV.14),
$\hat A_g$ takes the form:
$$\hat A_g=
< \prod_{a=1}^{2g-2} d\tilde\rho_a d\tilde{\bar
\rho_a} \prod_{I=1}^g d\alpha_I d\phi_I
\biggl|\prod_{j=1}^{g}\int dm^{U(1)}_j
\prod_{i=1}^{3g-3+N}
\delta(<M_T^i|b>)
\int_{a_j} dy_j v(y_j)\biggr|^2\eqno(IV.16)$$
$$
\biggl|\sum_{n_I=0}^{2g-2+N}\sum_{{a^+},a^-\in a}
\prod_{a^+=1}^{n_I}
[\exp (-\hp+2\phm) \hvem \epsilon^{klmn}(\dz x^{+\bar k}
\Gplb \Gpmb \Gpnb)](\tilde z_{a^+})$$
$$
\prod_{a^-=1}^{2g-2+N-n_I}
[\exp (-\hp+\php+\phm) \hvep\dz x^\ml \Gplb](\tilde z_{a^-})$$
$$
\xp\xm u
(e^\phm \hvem\lp)^{2N+g-1-n_I} (e^\php\hvep\lm)^{N+1-g+n_I}(w)
\biggr|^2
\prod_{r=1}^N
W_{G,r} (z_r,\bar z_r)>_{\tau},$$
where the location of $w$ is arbitrary.

Since there are no $e^{+\hm}$ terms and since the screening
charge of $\hm$ is now zero, none of the $e^{-\hm}$
terms can contribute. Also because there
are no extra $\hvepm$ fields, no unnecessary
$\psi^\pm$ fields can contribute to the
amplitude. The U(1) moduli, $\mUj$, is therefore
fixed to zero, and after performing the correlation functions
for the $u$, $v$, $\xi^\pm$, $\eta^\pm$,
$\phi^\pm$, $\hvepm$, and $\psi^\pm$ fields, one obtains

$$\hat A_g=
<\int
\prod_{a=1}^{2g-2} d\tilde\rho_a d\tilde{\bar
\rho_a} \prod_{I=1}^g d\alpha_I d\phi_I
\biggl|
\prod_{i=1}^{3g-3+N}
\delta(<M_T^i|b>)\biggr|^2
\eqno(IV.17)$$
$$\biggl|\sum_{n_I=0}^{2g-2+N}\sum_{{a^+},a^-\in a}
\prod_{a^+=1}^{n_I}
\epsilon^{klmn}(\dz x^{+\bar k}
\Gplb \Gpmb \Gpnb)(\tilde z_{a^+})
\prod_{a^-=1}^{2g-2+N-n_I}
\dz x^\ml \Gplb (\tilde z_{a^-})$$
$$d^{2g-2}
\prod_{r=1}^n[{p^+_r}]^{5\over 2}
\prod_{a^+=1}^{n_I} [\dsr(\tilde z_{a^+}) ]^{-{3\over 2}}
\prod_{a^-=1}^{2g-2+N-n_I} [\dsr(\tilde z_{a^-}) ]^{-{1\over 2}}
\biggr|^2
$$
$$
\prod_{r=1}^N
\int dz_r d\bar z_r |(p^+_r)^{-3}\exp(p^+_r \theta_r^\mlb
\Gml(z_r))|^2 \exp (ip^\mu_r x_\mu (z_r,\bar z_r)>_{\tau},$$
where the equations $\dsr (\tilde
z_a)=\lim_{z\to \tilde z_a}(\dz\rho) [E(z,\tilde z_a)]^{-1}$ and
$p^+_r=\lim_{z\to z_r}(\dz\rho) E(z,z_r)$ have been used, and
$$d\equiv{{\dz\rho(z)
\prod_{r=1}^N E(z,z_r)}\over
{(\sigma(z))^2\prod_{a=1}^{2g-2+N}E(z, \tilde z_a)}}
\eqno(IV.18)$$ is independent
of $z$ because it is a single-valued function with no zeroes or
poles on the surface ($E(y,z)$ and $\sigma (z)$ are defined in
equation (III.19)).

The correlation function for the $b$ and $c$ fields
is the same calculation as for the bosonic
string, and therefore can be obtained by comparing the known
covariant and light-cone bosonic string amplitudes.$^{39,40}$ Equivalence of
these bosonic amplitudes implies that
$$<|\prod_{i=1}^{3g-3+N}
\delta(<M_T^i|b>)\prod_{r=1}^N c(z_r)|^2>_{\tau}=
\det (Im\, \tau)|Z_1(\tau) d^{2-2g}\prod_{r=1}^n (p^+_r)^{-\half}
\prod_{a=1}^{2g-2+N} (\dsr)^{-\half}|^2.\eqno(IV.19)$$
Note that under rescalings
$p^+_r\to Cp^+_r$, this correlation function scales like $C^{3-3g-N}$
which cancels the rescaling of the light-cone moduli.

The correlation function for the $x^+$ and $x^-$ in $\hat A_g$
simply substitutes $Re(\rho)$ everywhere for $x^+$ and introduces
an extra factor of $$\det (Im\,\tau)^{-1} |Z_1|^{-2}
\eqno(IV.20)$$ from the
partition function.

After multiplying together equations (IV.17), (IV.19), and (IV.20),
$\hat A_g$ takes exactly the same form as $A_g^{LC}$
of equation (IV.10), but with $s^\pl$ and $s^\mlb$ replaced by
$\Gml$ and $\Gplb$.
Because these two sets of fields have
identical conformal weights, their correlation functions
are equivalent, implying that the truncated twistor-string scattering
amplitude, $\hat A_g$, agrees with the light-cone gauge
scattering amplitude, $A_g^{LC}$. Note that the conformal anomaly
contribution to the light-cone gauge amplitude is zero since
the contribution from the $x^\ml$ and $\dz x^\plb$ fields cancels
the contribution from the $s^\pl$ and $s^\mlb$ fields.

\vskip 24 pt
\centerline {\bf VI. Concluding Remarks}
\vskip 12 pt

In this paper, the gauge-fixed N=(2,0) twistor-string action was
used to calculate Type IIB Green-Schwarz superstring amplitudes with
an arbitrary number of loops and external massless states. The manifest
spacetime supersymmetry of these amplitudes gives them advantages
over superstring amplitudes calculated using the Neveu-Schwarz-Ramond
formalism. As was mentioned in the introduction, NSR amplitudes
contain divergences before summing over spin structures, giving rise
to ``multiloop ambiguities''$^{4,26}$ and complicating the analysis of
finiteness.$^{27}$ Furthermore, the ghost contributions to the fermionic
vertex operator$^3$ and the necessity of performing a GSO projection
makes scattering in fermionic backgrounds difficult to describe in
the NSR formulation of the superstring.

Since the Green-Schwarz
superstring scattering amplitudes derived in this paper do not suffer
from these problems, they may be useful in providing a better
understanding of the finiteness properties of superstrings, and in
allowing superstring scattering amplitudes to be calculated in
fermionic backgrounds.
Another possible application of the results in this paper is to find
new relations between theta-functions on higher-genus Riemann surfaces.
By comparing Green-Schwarz multiloop scattering amplitudes with their
NSR counterparts, one might discover generalizations of the well-known
Jacobi identity that relates theta-functions of different spin-structures
on the torus.$^1$

An obvious disadvantage of the scattering amplitude calculations
in this paper
is that they are manifestly invariant under only an SU(4)xU(1)
subgroup of the full SO(9,1) super-Poincar\'e transformations. One
possibility for covariantizing the amplitude calculations is to
quantize a recently proposed version of the twistor-string with
N=8 worldsheet supersymmetry.$^{41,42}$
Since this twistor-string version of the
Green-Schwarz superstring replaces all eight of the Siegel-symmetries
with worldsheet super-reparameterizations,
it might be possible to gauge-fix the
N=8 twistor-string action without breaking the SO(9,1)
super-Poincar\'e invariance. However, at this point, it is not known
how to gauge-fix the N=8 action to a free-field action, and even
though the N=8 twistor-string has been shown to be classically
equivalent to the ten-dimensional Green-Schwarz superstring, it is
unclear if the equivalence remains after quantization (similar statements
can be made about the N=1 and N=4 twistor-string versions of the
ten-dimensional Green-Schwarz superstring$^{43,44}$).

A second possibility for covariantizing the amplitude calculations is
to quantize the Lorentz-covariant version of the N=2 twistor-string
action without explicitly breaking the SO(9,1) super-Poincar\'e
invariance (at the present time, this possibility exists only
for the N=(2,0) twistor-string version of the heterotic Green-Schwarz
superstring, since Lorentz-covariant twistor-string actions are not
yet known for the non-heterotic superstrings).
It is likely that pure spinors (the definition of a pure spinor in
ten dimensions is a non-zero sixteen-component complex Weyl SO(9,1)
spinor, $\lambda^\alpha$, satisfying $\lambda^\alpha \gmu \lambda^\beta$
for $\mu$=0 to 9, where $\lambda^\alpha$ is defined up to a complex
projective transformation) would play a fundamental role
in any covariant quantization of the N=2 twistor-string since
breaking SO(9,1) down to SU(4)xU(1) is equivalent to selecting out
a pure spinor (the only non-zero component of this pure spinor is
the $1_{+1}$ component of the anti-chiral SO(8) spinor).

It is interesting that pure spinors arise naturally in the covariant
N=(2,0) twistor-string since
$D_\pm \Theta^\alpha \gmu D_\pm\Theta^\beta=0$
is implied by $D_\pm X^\mu=D_\pm\TgT$, which is an on-shell
equation of motion for the superfields. Using Howe's
observation that the on-shell supergravity and super-Yang-Mills classical
equations of motion are related to the existence of pure spinors in
loop superspace,$^{45}$ Tonin was able to show that the classical N=(2,0)
twistor-string action can be consistently coupled (i.e., coupled in
a manner that preserves the local gauge symmetries of the
two-dimensional action) to a supergravity and super-Yang-Mills
background if the background fields satisfy their classical equations
of motion.$^{19}$ If covariant quantization of the N=(2,0) twistor-string
were possible, one could see how the classical supergravity and
super-Yang-Mills equations of motion are modified by requiring the
full quantum action to be consistently coupled to the background fields.

\vskip 24pt
\centerline {\bf Appendix: Gauge-Fixing of the N=(2,0) Twistor-String}
\vskip 12pt
This appendix will review Section III.C of reference 17, in which the
Lorentz-covariant twistor-string action for the Green-Schwarz
heterotic superstring is gauge-fixed to a free-field action.
\vskip 12pt
A. The Lorentz-Covariant N=(2,0) Twistor-String Action
\vskip 12pt
The Lorentz-covariant action for the N=(2,0) twistor-string defined
on an N=(2,0) super-worldsheet with Minkowski metric is:$^{19,20,43}$
$$\int dz d\bar z d\ep d\em \{
-i(P_{\mu}^+ \hat\Pi_\ep^\mu -  P_\mu^- \hat\Pi_\em^\mu) +{1\over 2}
\hat\Phi^{+\bar q}\hat\Phi^{-q} \eqno(A.1)$$
$$-{1\over 2}\ep[\dzbar X_\mu (\hDp\TgT)-\hDp X_\mu (\dzbar \TgT)]
$$
$$
+{1\over 2}\pd
[\dzbar X_\mu (\hDm\TgT)-\hDm X_\mu (\dzbar \TgT)]\}$$
$$\hbox{with the chirality constraints, }\quad\hDm\hat\Phi^{+\bar q}=
\hDp\hat\Phi^{-q}=0,$$
$$\hbox{and the covariant derivatives, }\quad
\hat D_{\pm}\equiv\partial_{\kappa^\pm}+i\kappa^\mp[\dz
+e(z,\bar z) \dzbar +(\dzbar e(z,\bar z))M],$$
where  $[z,\kappa^+,\kappa^-,\bar z]$ are the coordinates for the
Minkowski-space worldsheet (note that
$\kappa^- =(\kappa^+)^*$, but $\bar z\not= z^*$),
$e(z,\bar z)$ is a real component field
independent of $\kappa^\pm$ and is the only remnant of the two-dimensional
super-vielbein ($\hdz \equiv -{i\over 2}
\{ \hDp,\hDm\}$=$\dz+e\dzbar+(\dzbar e)M$),
$M$ is the generator of two-dimensional
Lorentz rotations that measures the conformal weight with respect to
$\dzbar$ (i.e., $M$ commutes with everything except for
$[M,\dzbar]=\dzbar$ and $[M,\hat\Phi]={1\over 2}\hat\Phi$),
$\hat\Pi_{\kappa^\pm}^\mu\equiv\hat D_{\pm}X^\mu-i(
\hat D_{\pm}\TgT)$, $P^{\pm}_{\mu}$ are Lagrange multipliers for
$\hat\Pi_{\kappa^\pm}^\mu$,
and $X^\mu$, $\Theta^\alpha$, $\hat\Phi^{+\bar q}$,
$\hat\Phi^{-q}$ are N=(2,0) superfields whose $\kappa^\pm =0$ components
are the usual superspace variables of the Green-Schwarz heterotic
superstring, $x^\mu$, $\theta^\alpha$, $\phi^{+\bar q}$,
$\phi^{-q}$ for q=1 to 16.
Note that because the
Wess-Zumino term multiplying $\pp$ (or $\pd$) is chiral (or anti-chiral)
when $\hat\Pi^\mu_{\kappa^\pm}=0$, the action will be super-reparameterization
invariant after shifting $P^+_\mu$ and $P^-_\mu$ appropriately.
\footnote\dag{
It has recently been shown$^{41}$ that the second and third lines of
equation (A.1) can be written in a manifestly super-reparameterization
invariant form by introducing a superfield $E^\pm$ in the place of the
coordinate $\kappa^\pm$. Since this superfield can be gauge-fixed on-shell to
be proportional to $\kappa^\pm$, these two actions are classically equivalent,
where the proportionality constant is interpreted as the string tension.}

The equations of motion one gets from varying the unconstrained
superfields are:
$$\hDp\hat\Phi^{+\bar q}=\hDm\hat\Phi^{-q}=
\hat\Pi_{\kappa^\pm}=\pp\pd (\hDp(P_\mu^+ \Pi^\mu_{\bar z})+\hDm(P_\mu^-
\Pi^\mu_{\bar z})-
{i\over 2}(\hat\Phi^{+\bar q}\dzbar\hat\Phi^{-q}+\hat\Phi^{-q}
\dzbar\hat\Phi^{+\bar q}))\eqno(A.2)$$
$$=
\hDp P^{+\,\mu}+i\pp(\hDp\Theta^\alpha\gmu\dzbar\Theta^\beta)
-\hDm P^{-\,\mu}-i\pd(\hDm\Theta^\alpha\gmu\dzbar\Theta^\beta)$$
$$=(P_\mu^+ +{1\over 2}\pp\Pi_{\bar z\,\mu})(\gmu\hDp\Theta^\beta)
-(P_\mu^- +{1\over 2}\pd\Pi_{\bar z\,\mu})(\gmu\hDm\Theta^\beta)=0.$$
These equations imply that $\hDpm\Theta^\alpha\gmu\hDpm\Theta^\beta=0$
(i.e., $\hDpm\Theta^\alpha$ are ``pure spinors''), and that
$$\hDp P^{+\,\mu}
+{1\over 2}\Pi^\mu_{\bar z}+i\pp(\hDp\Theta^\alpha\gmu\dzbar\Theta^\beta)
=
\hDm P^{-\,\mu}
+{1\over 2}
\Pi^\mu_{\bar z}+i\pd(\hDm\Theta^\alpha\gmu\dzbar\Theta^\beta)$$
$$=A(\hDp\Theta^\alpha\gmu\hDm\Theta^\beta)
\eqno(A.3)$$
for some real N=2 superfield A. Therefore,
$$
\hat D_{\pm}(\Pi^\mu_{\bar z}-A
\hat\Pi^\mu_z)=(\hat\Pi^\mu_z)^2=\pp\pd[(\Pi^\mu_{\bar z}-
A\hat\Pi^\mu_z)^2+ {i\over 2}( \hat\Phi^{+\bar q}
\dzbar\hat\Phi^{-q}+\hat\Phi^{-q}\dzbar\hat\Phi^{+\bar q})]=0,
\eqno(A.4)$$
where $\hat\Pi^\mu_z\equiv\hdz X^\mu-i\hdz\TgT$=
$\hDp\Theta^\alpha\gmu\hDm\Theta^\beta$.
In addition to implying the usual superstring equations of motion for
the component fields, $x^\mu$, $\theta^\alpha$, and $\phi^p$,
$$\partial_-\phi^p=\partial_-\pi^\mu_+=\pi_{-\,\mu}(\gmu\partial_+
\theta^\beta)=(\pi_-^\mu)^2=(\pi_+^\mu)^2+i\phi^p\partial_+\phi^p=0,
\eqno(A.5)$$
where $\pi^\mu_\pm\equiv\partial_\pm
x^\mu-i(\partial_\pm \tgt )$,
$\partial_-\equiv\dz+e\dzbar$ and $\partial_+\equiv
(1-ae)\dzbar -a\dz$, these superfield equations fix the
values of the auxiliary fields in $X^\mu$, $\Theta^\alpha$,
$P^\pm_\mu$, $\Phi^{+ \bar q}$, and $\Phi^{-q}$.
\vskip 12pt
B. The Gauge-Fixing Procedure
\vskip 12pt
With the appropriate transformations of $P^\pm_\mu$,
$\hat\Phi^{+\bar q}$, and $\hat\Phi^{-q}$, the action of
equation (A.1) is invariant under the N=2 super-reparameterizations,
$$[\delta z=2R-\pp\hDp R -\pd\hDm R,~
 \delta\kappa^\pm=-i\hat D_{\mp} R,~
\delta \bar z=r+e\delta z]\eqno(A.6)$$
where $R(z,\pp,\pd,\bar z)$
is a real N=2 superfield and $r(z,\bar z)$ is a real
component field independent of $\kappa^\pm$
(from this super-reparameterization, $\delta\hDp =-i(\hDp\hDm R)
\hDp$
and $\delta\hDm=-i(\hDm\hDp R)\hDm$
where $\delta e=-\dz r -e\dzbar r+r \dzbar e$), under the six independent
$K_\beta$-transformations, $$[\delta\Theta^\alpha=(\hDp\Theta^\gamma
\gamma^\mu_{\gamma\delta}\hDm\Theta^\delta)(\gamma_\mu^
{\alpha\beta} K_\beta)-2
\hDp\Theta^\alpha (\hDm\Theta^\beta K_\beta)-2\hDm\Theta^\alpha
(\hDp\Theta^\beta K_\beta),$$
$$\delta X^\mu=i(\delta\Theta^\alpha
\gmu\Theta^\beta)]\eqno(A.7)$$
(only six are independent since
$\delta\Theta^\alpha\gmu\hDpm\Theta^\beta=0$ on-shell),
and under
the five independent complex $C^\alpha$-transformations,
$$[\delta\Theta^\alpha=\delta X^\mu=0, ~\delta P^{+\,\mu}
=\hDp C^\alpha\gmu\hDp\Theta^\beta,~
\delta\bar P^{-\,\mu}=\hDm \bar C^\alpha\gmu\hDm\Theta^\beta]\eqno(A.8)$$
(only five are independent since
$\delta P^\pm_\mu\gmu\hDpm\Theta^\beta=\hDpm\delta P^\pm_\mu=0$
on-shell).

In order to write the action in terms of free fields, it is necessary
to use the six $K_\beta$-transformations to gauge-fix to zero
$\gamma^+_{\dot a\beta}
\Theta^\beta$ for $\dot a$=1 to 6, and to use the five
$C^\alpha$-transformations to gauge-fix the non-auxiliary components
of $P^\pm_\mu$. Since none
of these gauge transformations involve derivatives on $K_\beta$
or $C^\alpha$, there are no propagating ghosts coming from this
gauge fixing. Furthermore, the N=(2,0)
super-reparameterizations, $r(z,\bar z)$ and
$R(z,\kappa^\pm,\bar z)$,
should be used to locally gauge-fix $e(z,\bar z)$ and
$A(z, \kappa^\pm,\bar z)$ to zero,
giving rise to the usual right and left-moving fermionic reparameterization
ghosts of conformal weight +2, two right-moving bosonic ghosts of
conformal weight +${3\over 2}$, and one right-moving fermionic ghost
of conformal weight +1.

Because six components of $\Theta^\alpha$ have been gauge-fixed
to zero, only an SU(4)xU(1) subgroup of the SO(9,1) Lorentz invariance
remains manifest in this N=(2,0) superconformal gauge.
Under this SU(4)xU(1) subgroup, the SO(8) anti-chiral spinor,
$(\gamma^+\Theta)^{\dot a}$, can be chosen to break up into
a $(1_{+1}, 6_0,1_{-1})$ representation, in which case
the SO(8) chiral spinor, $(\gamma^-\Theta)^a$, breaks up into a
$(4_{+{1\over 2}},\bar 4_{-{1\over 2}})$ representation,
and the SO(9,1) vector, $X^\mu$, breaks up into a $(1_0,1_0,4_{-{1\over 2}},
\bar 4_{+{1\over 2}})$ representation.

Since the constraint $\Dp\Theta^\alpha\gmu\Dp\Theta^\beta$=0
implies that $(\gamma^+\Dp\Theta)^{\dot a}(\gamma^+\Dp\Theta)^{\dot a}=0$,
it can be assumed that $\Dp[(\gamma^+\Theta)^{7}
-i(\gamma^+\Theta)^{ 8}]=0$
without loss of generality (if $
\Dp[(\gamma^+\Theta)^{ 7}
+i(\gamma^+\Theta)^{ 8}]=0$, simply exchange $\pp$
with $\pd$ everywhere). After making this choice, the constraints
$\Pi^\mu_{\kappa^\pm}$=0 can be used to combine the $X^\mu$ and
$\Theta^\alpha$
real superfields into the following chiral and anti-chiral complex
superfields:
$$\Psi^\pm\equiv (\gamma^+\Theta)^{ 7}
\pm i(\gamma^+\Theta)^{8},\quad
S^{+l}\equiv (\gamma^-\Theta)^l+i(\gamma^-\Theta)^{l+4},\quad
S^{-\bar l}\equiv (\gamma^-\Theta)^l-i(\gamma^-\Theta)^{l+4},$$
$$X^{+\bar l}\equiv X^l+iX^{l+4}+i \Psi^+ S^{-\bar l},
\quad X^{-l}\equiv X^l-iX^{l+4}+i\Psi^- S^{+l},
\quad X^\pm\equiv X^0\pm X^9,$$
$$\hbox{where  }\Dm\Psi^+=\Dm S^{+l}=\Dm X^{+\bar l}=\Dp\Psi^-=
\Dp S^{-\bar l}=\Dp X^{-l}=0,\eqno(A.9)$$
$$(\Psi^+)^*=\Psi^-,~ (S^{+l})^*=S^{-\bar l},~
(X^{+\bar l})^*=X^{-l},~(X^+)^*=X^+, ~(X^-)^*=X^-,$$
$$\hbox{and
$(\Psi^+,\Psi^-,S^{+l},S^{-\bar l},X^{+\bar l},X^{-l},X^+,X^-) $
transforms like a}$$
$$\hbox{ $(1_{+1},1_{-1},4_{+{1\over 2}},\bar 4_{-{1\over 2}},
\bar 4_{+{1\over 2}},4_{-{1\over 2}},1_0,1_0)$ representation of SU(4)xU(1)
for $l$=1 to 4.}$$

\vskip 12pt
C. The Gauge-Fixed Free-Field N=(2,0) Twistor-String Action
\vskip 12pt
In terms of these complex superfields, the action of equation (A.1)
in N=(2,0) superconformal gauge takes the following simple form:
$$S=\int dz d\bar z d\pp d\pd [{i\over 4}
(X^{+\bar l}\dzbar X^{-l}-X^{-l}\dzbar X^{+\bar l}) +
W^-\dzbar\Psi^+ -W^+\dzbar\Psi^- +{1\over 2}\Phi^{+\bar q}
\Phi^{-q}]\eqno(A.10)$$
with the constraints:
$$\Dm W^- -\Dm\Psi^-(X^-+iS^{+l}S^{-\bar l}) =
\Dp W^+ -\Dp\Psi^+(X^--iS^{+l}S^{-\bar l}) =\eqno(A.11)$$
$$\pp\pd[\dzbar X^{+\bar l}\dzbar X^{-l}-
\dzbar X^-\dzbar X^++{i\over 2}(\Phi^{+\bar q}\dzbar\Phi^{-q}+
\Phi^{-q}\dzbar\Phi^{+\bar q})]=$$
$$\Dp X^+-i\Psi^-\Dp\Psi^+=\Dm X^+-i\Psi^+\Dm\Psi^-=$$
$$\Dp X^--iS^{-\bar l}\Dp S^{+l}=
\Dm X^--iS^{+l}\Dm S^{-\bar l}=$$
$$\Dp X^{+\bar l}-2iS^{-\bar l}\Dp\Psi^+=\Dm X^{+\bar l}
=\Dp X^{-l}=\Dm X^{-l}-2iS^{+l}\Dm\Psi^-=
$$
$$\Dm\Psi^+=
\Dm S^{+l}=\Dm\Phi^{+\bar q}=\Dp\Psi^-=
\Dp S^{-\bar l}=\Dp\Phi^{-q}=0.$$
Note that the action is invariant under
$W^\pm \to W^\pm + D_{\pm} \Lambda^\pm$ and that the constraints
on $W^\pm$ are solved by $W^\pm=
\Psi^\pm (X^- \mp iS^{+l}S^{-\bar l})$.

It is easy
to check that the only effect of the right-moving constraints
on the superfields in the action is to fix their chiralities through
$$\Dm X^{+\bar l}=\Dp X^{-l}=\Dm\Psi^+=\Dp\Psi^-
=\Dm\Phi^{+\bar q}=\Dp
\Phi^{-q};\eqno(A.12)$$
to relate $W^+$ and $W^-$
through the condition
$$\Dp W^+ \Dm\Psi^- -\Dm W^- \Dp\Psi^++{
i\over 2}\Dp X^{+\bar l}\Dm X^{-l}=0;
\eqno(A.13)$$
and to require that $$\Dp\Psi^+\Dm\Psi^--i\Psi^+\dz\Psi^-
-i\Psi^-\dz\Psi^+=\dz X^+
\hbox{ for some real superfield $X^+$}.\eqno(A.14)$$ Furthermore,
the $\kappa^+=\kappa^-=0$ component of
$\dzbar X^+$ should equal the $\dzbar x^+$ component
field that appears in the
left-moving Virasoro constraint (this implies that $\int_C dz
d\ep d\em (\Psi^+ \Psi^-)=\int_C d\bar z (\dzbar x^+)$ for any closed
curve C).

\vskip 24pt
\centerline {\bf Acknowledgements}
\vskip 12pt
I would like to thank A. Galperin, S. Mandelstam, E. Melzer,
P. Pasti, F. Pezzella, M. Porrati,
A. Restuccia, M. Rocek, W. Siegel, J. Stephany,
M. Tonin, C. Vafa, P. van Nieuwenhuizen, and E. Verlinde
for useful discussions.
This work was supported by National Science Foundation grant
$\#$PHY89-08495.

\vskip 24pt

\centerline{\bf References}
\vskip 12pt

\item{(1)} Schwarz,J.H., Phys.Rep.89 (1982), p.223.

\item{(2)} Polyakov,A.M., Phys.Lett.B103 (1981), p.211.

\item{(3)} Friedan,D., Martinec,E., and Shenker,S., Nucl.Phys.B271
(1986), p.93.

\item{(4)} Verlinde,E. and Verlinde,H., Phys.Lett.B192 (1987), p.95.

\item{(5)} Green,M.B. and Schwarz,J.H., Nucl.Phys.B243 (1984), p.475.

\item{(6)} Mandelstam,S., Prog.Theor.Phys.Suppl.86 (1986), p.163.

\item{(7)} Restuccia,A. and Taylor,J.G., Phys.Rep.174 (1989), p.283.

\item{(8)} Carlip,S., Nucl.Phys.B284 (1987), p.365.

\item{(9)} Kallosh,R. and Morosov,A., Phys.Lett.B207 (1988), p.164.

\item{(10)} Gilbert,G. and Johnston,D., Phys.Lett.B205 (1988), p.273.

\item{(11)} Porrati,M. and van Nieuwenhuizen,P., Phys.Lett.B273 (1991), p.47.

\item{(12)} Mandelstam,S., ``Interacting-string picture of the
fermionic string'' in 1985 Santa Barbara Workshop on Unified
String Theories, eds. M.B. Green and D. Gross (World Scientific,
Singapore), p.577.

\item{(13)} Greensite,J. and Klinkhamer,F.R., Nucl.Phys.B291 (1987), p.557.

\item{(14)} Mandelstam,S., private communication.

\item{(15)} Green,M.B. and Schwarz,J.H., Nucl.Phys.B243 (1984), p.285.

\item{(16)} Siegel,W., Phys.Lett.B128 (1983), p.397.

\item{(17)} Berkovits,N., Nucl.Phys.B379 (1992), p.96.

\item{(18)} Penrose,R. and MacCallum,M.A.H., Phys.Rep.6C (1972), p.241.

\item{(19)} Tonin,M., Phys.Lett.B266 (1991), p.312.

\item{(20)} Ivanov,E.A. and Kapustnikov,A.A., Phys.Lett.B267 (1991), p.175.

\item{(21)} Sorokin,D.P., Tkach,V.I., Volkov,D.V., and Zheltukhin,A.A.,
Phys.Lett.B216 (1989), p.302.

\item{(22)} Banks,T., Dixon,L., Friedan,D., and Martinec,E.,
Nucl.Phys.B299 (1988), p.613.

\item{(23)} Friedan,D., private communication.

\item{(24)} Cohn,J., Nucl.Phys.B284 (1987), p.349.

\item{(25)} Melzer,E., J.Math.Phys.29 (1988), p.1555.

\item{(26)} Atick,J. and Sen,A., Nucl.Phys.B296 (1988), p.157.

\item{(27)} Mandelstam,S., ``The n-loop Amplitude: Explicit Formulas,
Finiteness, and Absence of Ambiguities'', preprint UCB-PTH-91/53,
October 1991.

\item{(28)} Ademollo,M., Brink,L., D'Adda,A., D'Auria,R.,
Napolitano,E., Sciuto,S., Del Giudice,E., DiVecchia,P., Ferrara,S.,
Gliozzi, F., Musto,R., Pettorini,R., and Schwarz,J., Nucl.Phys.B111
(1976), p.77.

\item{(29)} Rocek,M., private communication.

\item{(30)} Vafa,C., private communication.

\item{(31)} Verlinde,E. and Verlinde,H., Nucl.Phys.B288 (1987), p.357.

\item{(32)} Dijkgraaf,R., Verlinde,E., and Verlinde,H., Comm.Math.Phys.115
(1988), p.649.

\item{(33)} D'Hoker,E. and Phong,D.H., Comm.Math.Phys.125 (1989), p.469.

\item{(34)} Ooguri,H. and Vafa,C., Nucl.Phys.B361 (1991), p.469.

\item{(35)} Martinec,E., Phys.Lett.B171 (1986), p.189.

\item{(36)} Berkovits,N., Nucl.Phys.B304 (1988), p.537.

\item{(37)} Aoki,K., D'Hoker,E., and Phong,D.H., Nucl.Phys.B342 (1990),
p.149.

\item{(38)} Mandelstam,S., Nucl.Phys.B69 (1974), p.77.

\item{(39)} D'Hoker,E. and Giddings,S., Nucl.Phys.B291 (1987), p.90.

\item{(40)} Mandelstam,S., ``The interacting-string
picture and functional integration''
in 1985 Santa Barbara Workshop on Unified
String Theories, eds. M.B. Green and D. Gross (World Scientific,
Singapore), p.577.

\item{(41)} Delduc,F., Galperin,A., Howe,P., and Sokatchev,E.,
``A twistor formulation of the heterotic D=10 superstring with
manifest (8,0) worldsheet supersymmetry'', preprint BONN-HE-92-19,
JHU-TIPAC-920018, ENSLAPP-L-392-92, July 1992.

\item{(42)} Tonin,M., private communication.

\item{(43)} Berkovits,N., Phys.Lett.B232 (1989), p.184.

\item{(44)} Delduc,F., Ivanov,E., and Sokatchev,E., ``Twistor-like
superstrings with N=(1,0),(2,0),(4,0) worldsheet supersymmetry'',
preprint ENSLAPP-L-371-92, BOBB-HE-92-11, April, 1992.

\item{(45)} Howe,P.S., Phys.Lett.B258 (1991), p.141.

\end